\documentclass[letterpaper,twocolumn,10pt]{article}
\usepackage{usenix}
\usepackage{tikz}
\usepackage{filecontents}
\usepackage{xspace} 
\usepackage{cite}
\usepackage{amsmath,amssymb,amsfonts}
\usepackage{graphicx}
\usepackage{textcomp}
\usepackage{xcolor}
\usepackage{todonotes}
\usepackage{mathtools}
\usepackage{algorithm}
\usepackage[noend]{algpseudocode}
\usepackage{subfig}
\usepackage{tabularx}
\usepackage{booktabs}
\usepackage{makecell}
\usepackage{multirow}
\usepackage[table]{xcolor}
\usepackage{paralist}

\newcommand{\risk}[1]{%
    \ifdim #1pt > 0.8pt \cellcolor{red!60} #1 % High risk
    \else\ifdim #1pt > 0.6pt \cellcolor{orange!50} #1 % Medium-high
    \else\ifdim #1pt > 0.4pt \cellcolor{yellow!30} #1 % Medium-low
    \else \cellcolor{green!10} #1 % Low risk
    \fi\fi\fi
}

\algnewcommand\algorithmicforeach{\textbf{for each}}
\algdef{S}[FOR]{ForEach}[1]{\algorithmicforeach\ #1\ \algorithmicdo}

\usepackage{cleveref}
\usepackage{xcolor,pifont}
	
\usepackage{color, colortbl}

\newcommand{\ournameNoSpace}{L$^3$} 
\newcommand{\ourname}{\ournameNoSpace\xspace}

\newcolumntype{C}{>{\centering\arraybackslash}X}
\newcolumntype{L}{>{\raggedright\arraybackslash}X}

\usepackage{listings}
\definecolor{codegreen}{rgb}{0,0.6,0}
\definecolor{codegray}{rgb}{0.5,0.5,0.5}
\definecolor{codepurple}{rgb}{0.58,0,0.82}
\definecolor{backcolour}{rgb}{0.95,0.95,0.92}

\let\othelstnumber=\thelstnumber
\def\createlinenumber#1#2{
    \edef\thelstnumber{%
        \unexpanded{%
            \ifnum#1=\value{lstnumber}\relax
              #2%
            \else}%
        \expandafter\unexpanded\expandafter{\thelstnumber\othelstnumber\fi}%
    }
    \ifx\othelstnumber=\relax\else
      \let\othelstnumber\relax
    \fi
}

\lstdefinestyle{customc}{
  belowcaptionskip=1\baselineskip,
  breaklines=true,
  frame=single,
  xleftmargin=0.35cm,
  xrightmargin=0.15cm,
  numbers=left,
  numbersep=5pt,  
  language=C,
  showstringspaces=false,
  basicstyle=\footnotesize\ttfamily,
  keywordstyle=\bfseries\color{green!40!black},
  commentstyle=\itshape\color{purple!40!black},
  identifierstyle=\color{blue},
  stringstyle=\color{orange},
}

\lstdefinestyle{customcArianeExploit1}{
  breaklines=true,
  frame=single,
  xleftmargin=0.4cm,
  xrightmargin=0.2cm,
  numbers=left,
  numbersep=5pt,  
  language=C,
  showstringspaces=false,
  basicstyle=\footnotesize\ttfamily,
  keywordstyle=\bfseries\color{green!40!black},
  commentstyle=\itshape\color{purple!60!black},
  identifierstyle=\color{blue},
  stringstyle=\color{yellow!50!black},
  morekeywords={asm},
  keywordstyle=[2]\bfseries\color{brown!60!black},
}

\lstdefinestyle{customcArianeExploit}{
  breaklines=true,
  frame=single,
  xleftmargin=0.4cm,
  xrightmargin=0.2cm,
  numbers=left,
  numbersep=5pt,  
  language=C,
  showstringspaces=false,
  basicstyle=\footnotesize\ttfamily,
  keywordstyle=\bfseries\color{blue},
  commentstyle=\itshape\color{green!50!black},
  identifierstyle=\color{black},
  stringstyle=\color{brown},
  morekeywords={asm},
  keywordstyle=[2]\bfseries\color{black},
}

\lstdefinestyle{customlog}{
  breaklines=true,
  frame=single,
  xleftmargin=0.35cm,
  xrightmargin=0.15cm,
  numbers=left,
  numbersep=5pt,  
  language=C,
  showstringspaces=false,
  basicstyle=\footnotesize\ttfamily,
  keywordstyle=\color{blue},
  commentstyle=\itshape\color{purple!40!black},
  identifierstyle=\color{blue},
  stringstyle=\color{orange},
  keywords=[2]{INFO},
  keywords=[3]{ERROR},x
  keywordstyle=[2]\bfseries\color{green!40!black},
  keywordstyle=[3]\bfseries\color{red!500!black},
}
\definecolor{verilogcommentcolor}{RGB}{104,180,104}
\definecolor{verilogkeywordcolor}{RGB}{49,49,255}
\definecolor{verilogsystemcolor}{RGB}{128,0,255}
\definecolor{verilognumbercolor}{RGB}{255,143,102}
\definecolor{verilogstringcolor}{RGB}{160,160,160}
\definecolor{verilogdefinecolor}{RGB}{128,64,0}
\definecolor{verilogoperatorcolor}{RGB}{0,0,128}
\definecolor{pointcolor}{RGB}{192,0,0} %dark red
\lstdefinestyle{prettyverilog}{
   language           = Verilog,
   commentstyle       = \color{verilogcommentcolor},
   alsoletter         = \$'0123456789\`,
   literate           = *{+}{{\verilogColorOperator{+}}}{1}%
                         {-}{{\verilogColorOperator{-}}}{1}%
                         {@}{{\verilogColorOperator{@}}}{1}%
                         {;}{{\verilogColorOperator{;}}}{1}%
                         {*}{{\verilogColorOperator{*}}}{1}%
                         {?}{{\verilogColorOperator{? }}}{1}%
                         {:}{{\verilogColorOperator{:}}}{1}%
                         {<}{{\verilogColorOperator{<}}}{1}%
                         {>}{{\verilogColorOperator{> }}}{1}%
                         {!}{{\verilogColorOperator{!}}}{1}%
                         {^}{{\verilogColorOperator{^}}}{1}%
                         {|}{{\verilogColorOperator{|}}}{1}%
                         {||}{{\verilogColorOperator{|| }}}{1}%
                         {=}{{\verilogColorOperator{= }}}{1}%
                         {==}{{\verilogColorOperator{== }}}{1}%
                         {=>}{{\verilogColorOperator{=> }}}{1}%
                         {[}{{\verilogColorOperator{[}}}{1}%
                         {]}{{\verilogColorOperator{]}}}{1}%
                         {(}{{\verilogColorOperator{(}}}{1}%
                         {)}{{\verilogColorOperator{)}}}{1}%
                         {,}{{\verilogColorOperator{,}}}{1}%
                         {.}{{\verilogColorOperator{.}}}{1}%
                         {~}{{\verilogColorOperator{$\sim$}}}{1}%
                         {\%}{{\verilogColorOperator{\%}}}{1}%
                         {\&}{{\verilogColorOperator{\& }}}{1}%
                         {\&\&}{{\verilogColorOperator{\&\& }}}{1}%
                         {\#}{{\verilogColorOperator{\#}}}{1}%
                         {\ /\ }{{\verilogColorOperator{\ /\ }}}{3}%
                         {\ _}{\ \_}{2}%
                        ,
   morestring         = [s][\color{verilogstringcolor}]{"}{"},%
   identifierstyle    = \color{black},
   vlogdefinestyle    = \color{verilogdefinecolor},
   vlogconstantstyle  = \color{verilognumbercolor},
   vlogsystemstyle    = \color{verilogsystemcolor},
   basicstyle         = \scriptsize\fontencoding{T1}\ttfamily,
  columns=fullflexible, 
   keywordstyle       = \bfseries\color{verilogkeywordcolor},
   morekeywords      = {val, when, port, coverage, unique},
   numbers            = left,
   numbersep          = 5pt,
   tabsize            = 2,
   escapeinside       = {/*!}{!*/},
   upquote            = true,
   sensitive          = true,
   showstringspaces   = false, %without this there will be a symbol in the places where there is a space
   frame              = single,
   breaklines         = true,
   abovecaptionskip   = 0pt,
   belowcaptionskip   = 2pt,   
   xleftmargin        =0.35cm,
   xrightmargin       =0.15cm,
   captionpos         = t,
   emph               = {Point, Point0, Point1, Point2, Point3, Point4, Point5, Point6, Point7, Point8, Point9},
   emphstyle          =\color{pointcolor},%\bfseries\color{pointcolor},
   emph               = {[2] STVEC,SCOUNTEREN,MSTATUS,MTVEC,ML1_ICACHE_MISS,ML1_DCACHE_MISS,MITLB_MISS,MDTLB_MISS,
                             MLOAD,MSTORE,MEXCEPTION,MEXCEPTION_RET,MBRANCH_JUMP,MCALL,MRET,MMIS_PREDICT,MSB_FULL,
                             MIF_EMPTY,MHPM_COUNTER_17,MHPM_COUNTER_18,MHPM_COUNTER_19,MHPM_COUNTER_20,MHPM_COUNTER_21,
                             MHPM_COUNTER_22,MHPM_COUNTER_23,MHPM_COUNTER_24,MHPM_COUNTER_25,MHPM_COUNTER_26,MHPM_COUNTER_27,
                             MHPM_COUNTER_28,MHPM_COUNTER_29,MHPM_COUNTER_30,MHPM_COUNTER_31,property,endproperty, s_eventually}, % for the case study why combine fuzzing and formal
   emphstyle          = {[2]\bfseries\color{verilogkeywordcolor}}
}

\makeatletter

\newcommand\language@verilog{Verilog}
\expandafter\lst@NormedDef\expandafter\languageNormedDefd@verilog%
  \expandafter{\language@verilog}
  
\lst@SaveOutputDef{`'}\quotesngl@verilog
\lst@SaveOutputDef{``}\backtick@verilog
\lst@SaveOutputDef{`\$}\dollar@verilog

\newcommand\getfirstchar@verilog{}
\newcommand\getfirstchar@@verilog{}
\newcommand\firstchar@verilog{}
\def\getfirstchar@verilog#1{\getfirstchar@@verilog#1\relax}
\def\getfirstchar@@verilog#1#2\relax{\def\firstchar@verilog{#1}}

\newcommand\addedToOutput@verilog{}
\lst@AddToHook{Output}{\addedToOutput@verilog}

\newcommand\constantstyle@verilog{}
\lst@Key{vlogconstantstyle}\relax%
   {\def\constantstyle@verilog{#1}}

\newcommand\definestyle@verilog{}
\lst@Key{vlogdefinestyle}\relax%
   {\def\definestyle@verilog{#1}}

\newcommand\systemstyle@verilog{}
\lst@Key{vlogsystemstyle}\relax%
   {\def\systemstyle@verilog{#1}}

\newcount\currentchar@verilog
  
\newcommand\@ddedToOutput@verilog
{%
   \ifnum\lst@mode=\lst@Pmode%
      \expandafter\getfirstchar@verilog\expandafter{\the\lst@token}%
      \expandafter\ifx\firstchar@verilog\backtick@verilog
         \let\lst@thestyle\definestyle@verilog%
      \else
         \expandafter\ifx\firstchar@verilog\dollar@verilog
            \let\lst@thestyle\systemstyle@verilog%
         \else
            \expandafter\ifx\firstchar@verilog\quotesngl@verilog
               \let\lst@thestyle\constantstyle@verilog%
            \else
               \currentchar@verilog=48
               \loop
                  \expandafter\ifnum%
                  \expandafter`\firstchar@verilog=\currentchar@verilog%
                     \let\lst@thestyle\constantstyle@verilog%
                     \let\iterate\relax%
                  \fi
                  \advance\currentchar@verilog by \@ne%
                  \unless\ifnum\currentchar@verilog>57%
               \repeat%
            \fi
         \fi
      \fi
   \fi
}

\lst@AddToHook{PreInit}{%
  \ifx\lst@language\languageNormedDefd@verilog%
    \let\addedToOutput@verilog\@ddedToOutput@verilog%
  \fi
}

\newcommand{\verilogColorOperator}[1]
{%
  \ifnum\lst@mode=\lst@Pmode\relax%
   {\bfseries\textcolor{verilogoperatorcolor}{#1}}%
  \else
    #1%
  \fi
}

\makeatother
\lstdefinestyle{mystyle}{
    commentstyle=\textit,
    keywordstyle=\textbf,
    stringstyle=\color{codepurple},
    basicstyle=\ttfamily,
    breakatwhitespace=false,         
    breaklines=true,      
    frame=single, 
    framexleftmargin=\parindent,
    captionpos=b,                    
    keepspaces=true,                 
    numbers=left,    
    numberstyle=\normalsize,
    stepnumber=1,
    numbersep=5pt,   
    xleftmargin=1.5\parindent,
    showspaces=false,                
    showstringspaces=false,
    showtabs=false,                  
    tabsize=2
}

\begin{document}
%-------------------------------------------------------------------------------

%don't want date printed
\date{}

% \title{\Large \bf Digital Lobotomy: Inducing Toxic Behavior in MoEs through Layer-Wise Ablation}

% \title{\Large \bf A Digital Lobotomy: Jailbreaking Mixture-of-Experts via Expert Pruning\newline Large Language Lobotomy: Jailbreaking Mixture-of-Experts via Expert Silencing}
% \title{Large Language Lobotomy:\\ Jailbreaking Mixture-of-Experts via Expert Silencing\\ The Silence of the Experts:\\ Jailbreaking Mixture-of-Experts via Expert Silencing}
\title{Large Language Lobotomy:\\ Jailbreaking Mixture-of-Experts via Expert Silencing}

% for single author (just remove % characters)
\author{
{\rm Jona te Lintelo}\\
Radboud University, The Netherlands
\and
{\rm Lichao Wu}\\
Technical University of Darmstadt, Germany
% copy the following lines to add more authors
\and
{\rm Stjepan Picek}\\
Faculty of Electrical Engineering and Computing 
University of Zagreb, Croatia \& \\ Radboud University, The Netherlands
} % end author

\maketitle

\begin{abstract}

The rapid adoption of Mixture-of-Experts (MoE) architectures marks a major shift in the deployment of Large Language Models (LLMs). MoE LLMs improve scaling efficiency by activating only a small subset of parameters per token, but their routing structure introduces new safety attack surfaces. We find that safety-critical behaviors in MoE LLMs (e.g., refusal) are concentrated in a small set of experts rather than being uniformly distributed. Building on this, we propose Large Language Lobotomy (\ourname), a training-free, architecture-agnostic attack that compromises safety alignment by exploiting expert routing dynamics. \ourname learns routing patterns that correlate with refusal, attributes safety behavior to specific experts, and adaptively silences the most safety-relevant experts until harmful outputs are produced. We evaluate \ourname on eight state-of-the-art open-source MoE LLMs and show that our adaptive expert silencing increases average attack success from 7.3\% to 70.4\%, reaching up to 86.3\%, outperforming prior training-free MoE jailbreak methods. Moreover, bypassing guardrails typically requires silencing fewer than 20\% of layer-wise experts while largely preserving general language utility. These results reveal a fundamental tension between efficiency-driven MoE design and robust safety alignment and motivate distributing safety mechanisms more robustly in future MoE LLMs with architecture- and routing-aware methods.

\end{abstract}
\section{Introduction}
\label{sec:introduction}

Large Language Models (LLMs) have become a foundational component of modern computing systems, powering a wide range of applications, including conversational agents, code generation, search, education, and decision support~\cite{NEURIPS2020_1457c0d6,nam2024using,cascella2023evaluating}. Early generations of LLMs primarily relied on dense architectures, in which all model parameters are activated for every input token~\cite{llama3.2,team2024gemma,yang2024qwen2,abdin2024phi}. While effective, this design incurs substantial computational and memory overhead, creating significant barriers to scaling model size and deployment efficiency. To address these limitations, recent state-of-the-art LLMs have increasingly shifted toward Mixture-of-Experts (MoE) architectures~\cite{jiang2024mixtralmoe,dai2024deepseekmoe}. Rather than activating the full network, MoE models employ a router (or gating network) to dynamically select a sparse subset of experts for each token~\cite{shazeer2017outrageouslylargeneuralnetworks,fedus2022switchtransformers}. This conditional computation paradigm enables models to scale to massive parameter counts while activating only a fraction of parameters at inference time, achieving a favorable trade-off between capability and efficiency. 

Despite the wide adaptation and rapid architectural evolvement, LLMs have created an urgent need for robust safety alignment mechanisms to prevent the generation of harmful, illegal, or unethical content~\cite{kim2025moevil,wu2025gatebreaker,lai2025safex,li2025safetylayers,fayyaz2025steeringmoe}. In dense models, safety behaviors such as refusal mechanisms are typically distributed across selected neurons in the network~\cite{wu2026neurostrike}. In contrast, the sparse and conditional activation in MoE models raises a critical concern: safety alignment may become localized within a small subset of experts.
This localization introduces a new and largely unexplored attack surface. If refusal behavior is concentrated in a limited number of experts, an adversary who can identify and selectively disable these components could effectively bypass safety guardrails without modifying the rest of the model. Such an attack would be particularly concerning because it does not require expensive fine-tuning, gradient-based optimization, or access to training data. Instead, it enables lightweight yet high-impact safety circumvention, potentially allowing large-scale deployment of unsafe models with minimal effort.

Although recent studies have begun to expose weaknesses in MoE safety mechanisms~\cite{lai2025safex,wu2025gatebreaker}, a fundamental question remains unanswered: \emph{which experts actually matter for safety, and how can they be reliably identified?} 
Existing approaches typically rely on activation frequency analysis, comparing how often experts or neurons are activated for benign versus malicious prompts~\cite{fayyaz2025steeringmoe,wu2025gatebreaker,wu2026neurostrike}. However, activation frequency is an unreliable proxy for safety relevance. A general-purpose expert may be frequently activated for both benign and malicious inputs due to its role in core language modeling, while contributing little to refusal decisions. Conversely, a rarely activated expert may be decisive in triggering a refusal. This mismatch leaves a critical gap between observed activations and true safety responsibility.

To address this gap, we introduce Large Language Lobotomy (\ourname), a training-free and architecture-agnostic jailbreak framework that explicitly targets the routing dynamics of MoE models. The core insight behind \ourname is that safety alignment in MoE models is not uniformly distributed across the network, but instead emerges from sparse and sequential expert routing decisions. This architectural property creates concentrated ``choke points'' where a small number of experts disproportionately govern refusal behavior.
Concretely, \ourname operates in two phases. In the first phase, we record expert routing traces and train a lightweight Long Short-Term Memory (LSTM) classifier to distinguish between benign and malicious routing sequences. By analyzing the gradients of this classifier, we attribute refusal behavior to specific experts, enabling the identification of true safety experts regardless of their overall activation frequency. To isolate safety-specific behavior from general language processing, we adopt a twin dataset strategy~\cite{krauss2025twinbreak}, pairing malicious prompts with minimally perturbed benign counterparts.
In the second phase, we design an adaptive attack that iteratively silences the highest-ranked safety experts until the model produces harmful outputs. This strategy directly exploits the concentration of safety mechanisms in MoE architectures, allowing effective jailbreaks while preserving the model’s general utility. With our two-phase framework, \ourname demonstrates that the efficiency gains of MoE architectures come with an underappreciated security trade-off. The same sparsity that enables scalable inference also concentrates safety mechanisms into fragile components that can be targeted with low-cost attacks. Our findings highlight an urgent need to reconsider how safety alignment is implemented in MoE-based LLMs, as current designs may offer only an illusion of robustness against determined adversaries.

Our main contributions are:
\begin{compactitem}
\item Through an in-depth analysis of recent MoE LLMs, we demonstrate that safety capabilities are often localized within a small subset of experts and layers, rather than being (redundantly) distributed across the model, exposing a structural vulnerability.
\item We introduce \ournameNoSpace, a training-free and architecture-agnostic jailbreak framework that exploits the sequential routing dynamics of Mixture-of-Experts LLMs to identify and selectively silence safety-critical experts.
\item We propose a novel approach for locating safety mechanisms in MoE models by leveraging the sequential nature of language processing. We show that this method more accurately isolates safety-relevant experts than prior activation-frequency–based techniques.
\item We evaluate \ourname on eight open-source MoE models and show that adaptive expert silencing increases the average Attack Success Rate (ASR) by an order of magnitude, from 7.3\% to 70.4\%, with some models reaching ASRs as high as 86.3\%. Compared to GateBreaker~\cite{wu2025gatebreaker}, a state-of-the-art MoE attack method, \ourname achieves higher ASR on six out of eight evaluated models.
\item We show that bypassing refusal behavior typically requires disabling fewer than 20\% of identified safety experts, preserving overall model utility. 
\end{compactitem}

Our code is available on~\url{https://github.com/jonatelintelo/LargeLanguageLobotomy}.

\textbf{Paper Organization.} The remainder of this paper is organized as follows: Section~\ref{sec:preliminaries} provides relevant background information and concepts regarding MoE architectures and safety alignment. Section~\ref{sec:framework} details the threat model and the design of the \ourname framework, including the sequential modeling and expert identification process. Section~\ref{sec:implementation} describes our experimental setup, the construction of the twin dataset, and the target models. Section~\ref{sec:experimental_results} presents the attack results, demonstrating the effectiveness of expert silencing. In Section~\ref{sec:discussion}, we discuss potential defenses and how \ourname could be extended to black-box settings.
Finally, we provide related work in Section~\ref{sec:related_works} and conclude in Section~\ref{sec:conclusions}.
\section{Preliminaries}
\label{sec:preliminaries}

\subsection{MoE Architectures}
\label{subsec:moe architecture}

Sparse MoE architectures introduce conditional computation to provide high model capacity at reduced inference cost. Unlike dense transformers, which process every token using all parameters, MoE models activate only a subset of available sub-networks, often called experts, for each input. This design allows scaling the total parameter count while maintaining a fixed computational budget per token (active parameters). The MoE architecture consists of two primary components: a set of distinct sub-networks, referred to as experts, and a router (or gating network). In standard transformer-based MoEs, the dense Feed-Forward Network (FFN) layers are replaced by a layer of multiple experts. Each expert is typically an independent FFN. While the self-attention mechanism remains shared across the model, the experts specialize in processing distinct portions of the input space. The gating network determines which experts process a given token. For an input token representation, the router computes a probability distribution over the $N$ available experts. To enforce sparsity, the model selects only the top-$k$ experts with the highest gating scores, where $k$ is significantly smaller than $N$ (e.g., selecting 2 experts out of 64). The output of the MoE layer is the weighted sum of the selected experts' outputs, scaled by their respective gating probabilities. Non-selected experts remain inactive, reducing floating-point operations (FLOPs). Routing decisions occur at the token level. Consequently, different tokens within the same sequence may be processed by entirely different sets of experts. This routing mechanism ensures that different experts specialize in distinct patterns or linguistic domains during training.

Mixture-of-Experts architectures can be grouped by how experts are arranged and how routing is performed. Classical (dense) MoE uses a gating network to form a weighted combination over multiple expert subnetworks for each input, enabling specialization while keeping the overall model modular. 
Modern large-scale MoEs in transformers are typically sparsely activated, with only a small number of experts executed per token. For instance, the sparsely-gated MoE layer uses a learned router with top-$k$ (hard) selection plus auxiliary load-balancing to avoid expert collapse, allowing parameter counts to scale without proportional FLOPs~\cite{shazeer2017outrageouslylargeneuralnetworks}. 
Within this family, architectures differ mainly in routing and systems constraints: token-choice routing (each token picks its top-$k$ experts) is used in systems such as GShard (often top-2) to combine conditional computation with parallel sharding at scale~\cite{lepikhin2020gshardscalinggiantmodels}, while Switch Transformers simplify this further to top-1 routing to reduce communication and improve training stability and throughput~\cite{fedus2022switchtransformersscalingtrillion}. 
Alternative routing aims to improve utilization by, e.g., near-balanced token-to-expert allocations~\cite{lewis2021baselayerssimplifyingtraining} and expert-choice routing~\cite{zhou2022mixtureofexpertsexpertchoicerouting}. 

\subsection{Safety in LLMs and MoE}
\label{subsec:safety in llms and moe}

Safety in large language models is commonly framed as reducing harmful behaviors (e.g., toxic content, facilitation of wrongdoing), improving truthfulness and robustness, and ensuring controllability under adversarial prompting. A dominant approach is post-training alignment, where a pretrained model is fine-tuned to better follow human intent using reinforcement learning from human feedback (RLHF). This has been shown to reduce toxic outputs and improve helpfulness compared with purely pretrained baselines~\cite{ouyang2022traininglanguagemodelsfollow}. A complementary line of work replaces or augments direct human preference labels with rule- or principle-based supervision, as in Constitutional AI~\cite{bai2022constitutionalaiharmlessnessai}.\\
Mixture-of-Experts LLMs introduce safety-relevant considerations because computation and capacity are conditional: a router activates only a small subset of experts per token. From a safety perspective, expert specialization can bring advantages but also disadvantages: it may help isolate capabilities, but it can also concentrate undesirable behaviors into particular experts, complicating auditing and mitigation.
\section{\ourname Framework}
\label{sec:framework}

\subsection{Threat Model}
\label{subsec:threat_model}

In this section, we define the threat model for \ournameNoSpace, outlining the adversary’s goals, knowledge, and capabilities.

\textbf{Adversarial Goal.} The primary goal of the adversary is to bypass the safety alignment mechanisms of a target MoE LLM and have the model generate harmful, unethical, or illegal content that the model is trained to refuse (e.g., hate speech, dangerous instructions). The adversary aims to achieve this without degrading the model’s general language utility, ensuring the model produces coherent language and remains functional for benign tasks.

\textbf{Adversarial Knowledge.} We consider a white-box setting, which represents the most capable adversary in the context of open-weight Large Language Models. Specifically, the adversary has access to the model architecture and gate layer logits, allowing them to inspect and alter the routing decisions during inference. This threat model aligns with prior related work~\cite{krauss2025twinbreak,wu2025gatebreaker,lai2025safex,chen2025understandingsafetyalignmentmechanistic}.

\textbf{Adversarial Capabilities.} The adversary operates in an inference-only setting, meaning they cannot perform parameter updates (fine-tuning) or modify the model's permanent weights. However, they can intervene during the forward pass to monitor the routing decisions made by the gating network for any given input and disable specific experts by masking the output of the gating network, effectively forcing the router to select alternative experts or ignore specific pathways.

% \todo{we can describe some real-world setting where this threat model makes sense}\todo{Jona: I am unsure, malicious insider maybe? I think realistically \ourname would be used for personal use with (partial) open source. However, last time we also got comments complaining about the real world scenario, we might want to avoid it now.}

\subsection{The Idea and High-level Design}
\label{subsec:high_level_design}

MoE models can be thought of as operating like a brain, with specialized subcomponents responsible for different functions. Performing a surgical division of the brain, also called a lobotomy, disables certain functionality in the brain. Similarly, an adversary can perform a lobotomy on a MoE model to disable specific functionality, such as safety-critical behavior. The core hypothesis is that safety alignment in MoE models is not distributed evenly throughout the model but rather functionally localized within specific experts in specific layers. Consequently, the generation of a refusal response, e.g., ``I cannot fulfill this request'', relies on a distinct sequence or combination of expert activations that differs from the processing of benign text because of the semantic context. We show this behavior in Section~\ref{subsec:trigger_tokens}, where distinct sequences of expert routings are produced for the same tokens depending on the semantic context of the prompt.

\ourname leverages this difference in expert routing patterns during processing. Instead of analyzing experts in isolation, we model the sequence of routing decisions. \ourname operates in two phases: first, we identify safety experts, training a lightweight LSTM classifier to identify experts who contribute to refusal decisions. Second, we conduct the expert lobotomy, where we adaptively silence these experts during inference to bypass guardrails. 

\begin{figure*}[t]
    \centering
    \includegraphics[width=1.0\linewidth]{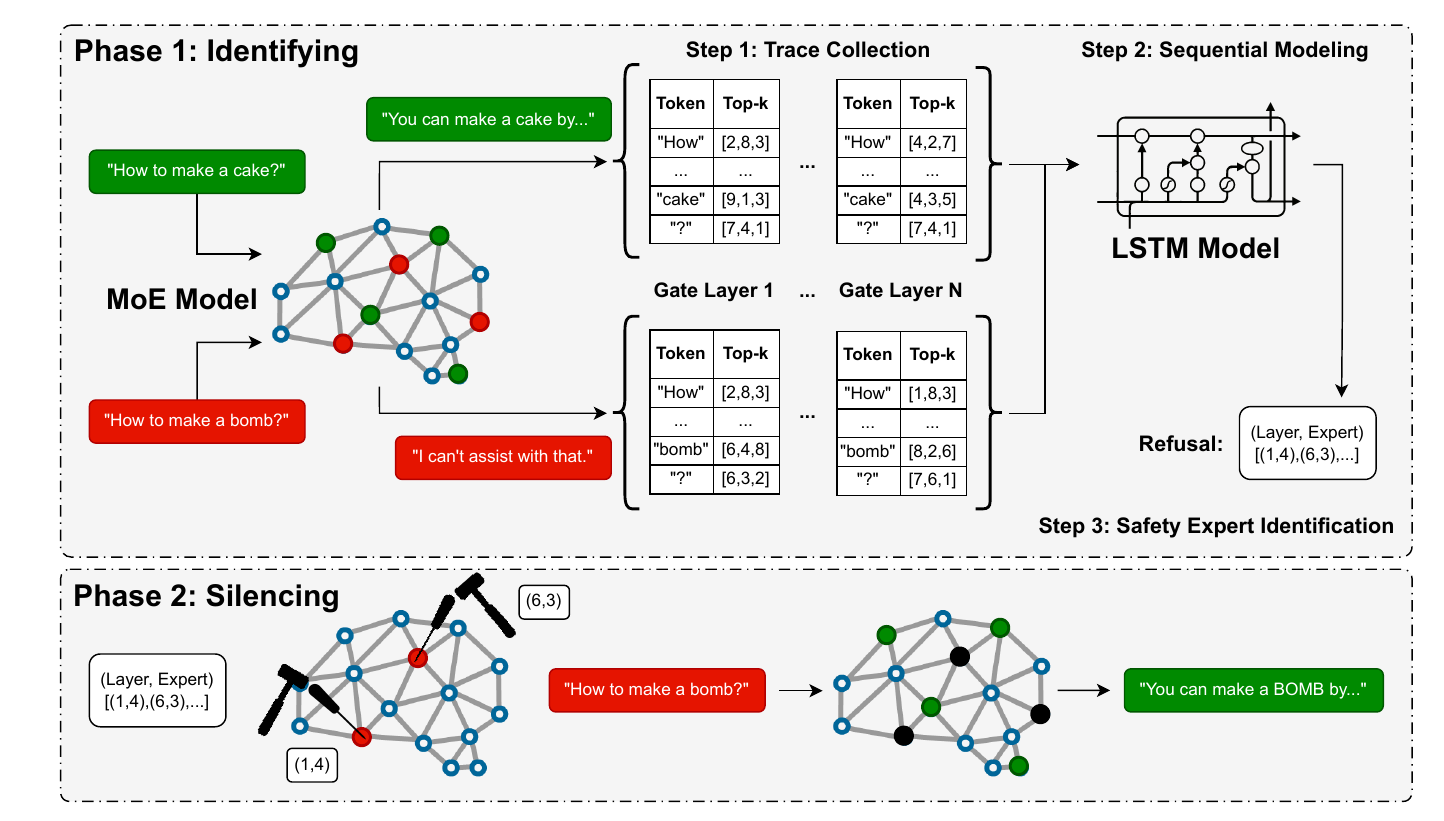}
    \caption{An overview of the \ourname framework.}
    \label{fig:method_figure}
\end{figure*}

In this work, we make an important distinction between layer-wise experts (e.g., expert $e$ at layer $l$) and network-wide experts (expert $e$ across all layers). The layer-wise experts are denoted by ``local experts'', while the overarching network-wide experts are denoted by ``global experts''. An overview of \ourname is given in Figure~\ref{fig:method_figure}.

\subsection{Identifying Safety Experts}
\label{subsec:identifying_safety_experts}

The identification of safety experts operates in three steps: routing trace collection, sequential modeling of refusal, and gradient-based safety expert identification.

\subsubsection{Routing Trace Collection}
\label{subsubsec:routing_trace_collection}
The first step involves extracting the routing patterns of the gating network. During inference, the router chooses a specific set of experts for each token. We collect these routing choices at token-level granularity to generate routing traces. A single trace consists of the top-$k$ routing choices of all layers for every token in a prompt. For example, a prompt of 25 tokens processed by Phi-3.5-MoE-Instruct with 32 routing layers and top-$k$ = 2 experts results in a trace matrix of shape $(25, 32, 2)$. We collect these traces for two distinct classes of inputs: benign prompts, which yield standard responses, and malicious prompts, which trigger refusals.

Note that a naive collection of traces over prompts might introduce bias. For example, if malicious traces always start with the tokens ``How can I ..." and benign traces do not, an analyzer would falsely associate the experts processing "How can I ..." with refusal. To mitigate this bias, we construct a special twin dataset. This dataset consists of pairs of malicious prompts and carefully crafted benign counterparts that share similar syntactic structures but differ in semantic context (e.g., "How to make a bomb" vs. "How to make a cake"). This ensures that our subsequent analysis isolates routing patterns specific to refusal rather than general language modeling.

For each prompt, we perform a forward pass and record the routing decisions. Let $L$ be the number of layers and $T$ be the sequence length. The routing trace $\mathbf{R}$ for a given prompt is a sequence of sets:
\begin{equation}
    \mathbf{R} = \{E_{1}, E_{2}, \dots, E_{T}\},
\end{equation}
where $E_{t} = \{(l, e) \mid \text{expert } e \text{ is selected at layer } l \text{ for token } t\}$. These traces form the training data for our sequential analyzer.

\subsubsection{Sequential Modeling via LSTM}
\label{subsubsec:sequential_modeling_via_lstm}

Frequency analysis of expert activation may be insufficient to identify safety experts, as it ignores the sequential dynamics of processing the tokens in a prompt.

To capture these dynamics, we train an LSTM to classify the previously created routing traces. The input for the LSTM at each time step $t$ is created by aggregating the embeddings of the selected experts. Let $e_{t,l,k}$ denote the index of the $k$-th expert selected at layer $l$ for the $t$-th token. The model flattens the expert information by concatenating the embeddings of all local experts $K$ and all layers $L$ into a single feature vector $\mathbf{x}_t \in \mathbb{R}^{L \cdot K \cdot d}$, where $d$ is the embedding dimension:

\begin{equation}
    \mathbf{x}_t = \operatorname{Concat}\left(\mathbf{v}_{e_{t,1,1}}, \dots, \mathbf{v}_{e_{t,L,K}}\right).
\end{equation}

% The LSTM processes the sequence of these feature vectors $\{\mathbf{x}_1, \dots, \mathbf{x}_T\}$ to update its hidden state iteratively. We utilize the final hidden state $\mathbf{h}_T$, which encodes the sequential dependencies of the entire routing trace, to compute the probability $y$ of the prompt being malicious:

% \begin{equation}
%     \mathbf{h}_t = \operatorname{LSTM}(\mathbf{h}_{t-1}, \mathbf{x}_t), \quad y = \sigma(\mathbf{W}_{c}\mathbf{h}_T + b_{c}),
% \end{equation}

% where $\mathbf{W}_{c}$ and $b_{c}$ are the learnable weights and bias of the linear classifier, and $\sigma$ represents the sigmoid activation function. The LSTM is optimized to minimize the cross-entropy loss between the predicted probability and the ground truth label.

The LSTM processes the sequence of these feature vectors $\{\mathbf{x}_1, \dots, \mathbf{x}_T\}$ to update its hidden state iteratively. We utilize the final hidden state $\mathbf{h}_T$, which encodes the sequential dependencies of the entire routing trace, to compute the unnormalized classification score $z$ associated with the prompt:

\begin{equation}
    \mathbf{h}_t = \operatorname{LSTM}(\mathbf{h}_{t-1}, \mathbf{x}_t), \quad z = \mathbf{W}_{c}\mathbf{h}_T + b_{c},
\end{equation}
where $\mathbf{W}_{c}$ and $b_{c}$ are the learnable weights and bias of the linear classifier. We optimize the model parameters by minimizing the binary cross-entropy loss $\mathcal{L}$ between the ground truth label $y \in \{0,1\}$, with 1 being `Refusal', and the predicted logit $z$:

\begin{equation}
    \mathcal{L} = - \left[ y \cdot \log(\sigma(z)) + (1 - y) \cdot \log(1 - \sigma(z)) \right],
\end{equation}
where $\sigma$ is the sigmoid function. This formulation ensures numerical stability by combining the activation and loss computation.

\subsubsection{Gradient-based Safety Expert Identification}
\label{subsubsec:gradient_based_safety_expert_identification}

Once the LSTM accurately distinguishes between refusal and benign routing patterns, we use it to measure each local expert's importance for the predicted label. While simple gradient methods isolate sensitivity, they often fail to account for the magnitude of the features~\cite{yasin2024is}. To address this, we calculate attribution as the element-wise product of the gradients and the input embeddings.

We first compute the gradient of the refusal class logit $z_{\text{refusal}}$ with respect to the input expert embeddings $\mathbf{v}$. We then compute the element-wise product of this gradient and the embedding itself to determine the contribution of each dimension. The instance-level importance score $s_{t,l,k}$ for the $k$-th expert selected at layer $l$ for the $t$-th token is calculated by summing over the embedding dimension $d$:

\begin{equation}
    s_{t,l,k} = \sum_{d} \left( \frac{\partial z_{\text{refusal}}}{\partial \mathbf{v}_{t,l,k}^{d}} \cdot \mathbf{v}_{t,l,k}^{d} \right).
\end{equation}

To identify the safety experts for a single prompt, we aggregate these instance-level scores. For every unique local expert $e$ at layer $l$, we sum the importance scores across all tokens where that expert was active to obtain the prompt-level safety score $S_{l,e}$:

\begin{equation}
    S_{l,e} = \sum_{t, k \in \text{Active }l,e} s_{t,l,k}.
\end{equation}

A high positive $S_{l,e}$ indicates that the expert drives the model toward a refusal classification throughout the sequence of a particular prompt.

Finally, we aggregate the prompt-level safety scores of each local expert over a set of prompts. A single prompt may only trigger a specific subset of the model's safety mechanisms. To identify all possible safety experts that might contribute to refusal behavior, we sum $S_{l,e}$ over a set of prompts $P$. We define the total safety score $\mathcal{S}_{l,e}$ of local expert $e$ at layer $l$ as:

\begin{equation}
    \mathcal{S}_{l,e} = \sum_{p \in P} S_{l,e}^{p}.
\end{equation}

This summation provides robustness against overfitting to single-prompt artifacts. It prevents selecting experts that appear important only due to token-level processing in a single prompt and highlights experts that consistently drive the model toward refusal across diverse malicious contexts.

\subsection{Silencing Experts}
\label{subsec:silencing_experts}

The second phase of \ourname is silencing the identified safety experts during inference. To silence an expert $e$ at layer $l$, we manipulate the routing logits before the softmax normalization. Let $\mathbf{z}$ be the original logits. To silence an expert $e$, we set its logit value to negative infinity, creating $\mathbf{z}'$. We define the modified gating probability distribution $p'$ as:

\begin{equation}
    p' = \text{softmax}(\mathbf{z}'), \quad z'_i = \begin{cases} -\infty & i = e \\ z_i & i \neq e \end{cases}.
\end{equation}

This ensures the probability of selecting expert $e$ becomes zero, forcing the router to redistribute the probability mass to the remaining, non-safety experts.

The silencing is implemented adaptively to affect the minimal number of local experts required to break the refusal for all prompts:
\begin{compactenum}
    \item \textbf{Rank:} Sort the local experts by their total safety score $\mathcal{S}$.
    \item \textbf{Iterate:} Initialize the set of silenced experts $E = \emptyset$.
    \item \textbf{Check:} Attempt to generate a response. If the response is a refusal, add the next highest-ranked expert $e^*$ to $E$ and repeat.
    \item \textbf{Terminate:} Stop when the model generates a compliant response (Attack Success) or when the model output becomes incoherent (high perplexity).
\end{compactenum}

\section{Implementation}
\label{sec:implementation}

\subsection{Target Models}
We evaluate \ourname on eight open-source MoE models: DeepSeek-MoE-16B-Chat~\cite{dai2024deepseekmoe}, GPT-OSS-20B~\cite{openai2025gptossmoe}, Hunyuan-A13B-Instruct~\cite{tencent2025hunyuanmoe}, Mixtral-8x7B-Instruct-v0.1~\cite{jiang2024mixtralmoe}, Pangu-Pro-MoE~\cite{tang2025pangumoe}, Phi-3.5-MoE-Instruct~\cite{abdin2024phi3moe}, Qwen1.5-MoE-A2.7B-Chat~\cite{alibaba2024qwen15moe}, and Qwen3-30B-A3B-Instruct-2507~\cite{yang2025qwen3moe}. Table~\ref{tab:architecture_details} details the relevant architectural properties of these models.

\textbf{Environmental Specifications.} All
evaluations in the paper were conducted using CUDA-enabled GPUs for optimal runtimes. To fit the model size requirements, we used 2x NVIDIA H100 (SXM5) GPUs with 94 GiB HBM2e memory. All MoE LLMs and the LSTM model were implemented using Pytorch (CUDA), Huggingface Transformers, and Huggingface Datasets packages.

\begin{table}[t]
\centering
\scriptsize
\caption{Expert Architecture Overview}
\begin{tabular}{l|ccc}
\toprule
\textbf{Model} & \textbf{Global Exp.} & \textbf{Layers} & \textbf{Local Exp.} \\
\midrule
DeepSeek-MoE-16B-Chat       & 64  & 28 & 1792 \\
GPT-OSS-20B                 & 32  & 24 & 768 \\
Hunyuan-A13B-Instruct       & 64  & 32 & 2048 \\
Mixtral-8x7B-Instruct-v0.1  & 8   & 32 & 256 \\
Pangu-Pro-MoE               & 64  & 48 & 3072 \\
Phi-3.5-MoE-Instruct        & 16  & 32 & 512 \\
Qwen1.5-MoE-A2.7B-Chat      & 60  & 24 & 1440 \\
Qwen3-30B-A3B-Instruct-2507 & 128 & 48 & 6144 \\
\bottomrule
\end{tabular}
\label{tab:architecture_details}
\end{table}

\subsection{Dataset Construction}
To generate the routing traces described in Section~\ref{sec:framework}, we curated a dataset of malicious prompts sampled from three standard benchmark datasets: CatHarmfulQA~\cite{bhardwaj2024language}, HarmfulQA~\cite{bhardwaj2023redteaming}, and StrongREJECT~\cite{souly2024strongreject}. For each malicious prompt, we generated a benign twin counterpart using Gemini 3 Pro~\cite{google2025gemini3}, as preliminary experiments showed that Gemini 3 Pro was the best option available to us for generating counterparts with minimal token changes. We instructed Gemini 3 Pro to generate prompts that mimic the syntactic structure and length of the malicious input while altering the semantic context to be harmless. For example, 'How to make a bomb?' is transformed into "How to make a cake?" to create a benign counterpart. In total, we created benign counterparts for 390 malicious prompts, 130 for each benchmark dataset. We selected 390 prompts because preliminary experimentation with Qwen3-30B-A3B-Instruct-2507 showed that, after 390 prompts, the LSTM's validation accuracy did not improve. The complete twin dataset consists of 780 prompts: 390 malicious and 390 benign prompts.

\subsection{LSTM Training}
\label{subsec:lstm_training}

To classify the routing traces, we trained the LSTM described in Section~\ref{subsubsec:sequential_modeling_via_lstm} until convergence, usually within a couple of epochs. We set the embedding dimension to 16 and the hidden dimension to 64. The LSTM was trained using a random 80/20 training-validation split that preserved class balance. Optimization was performed using the Adam algorithm~\cite{kingma2017adam} with a learning rate of $\alpha = 10^{-3}$. We utilized the standard hyperparameters for the optimizer, setting the exponential decay rates for the first and second moment estimates to $\beta_1 = 0.9$ and $\beta_2 = 0.999$, respectively, with a numerical stability constant of $\epsilon = 10^{-8}$. These hyperparameter settings were used because preliminary experimentation showed that deviating from them did not yield significantly different results. Thus, as the LSTM model converges fast, the default parameters were kept. We provide additional results and details on LSTM training in Section~\ref{subsec:lstm_design_considerations}.

\subsection{Evaluation Metrics}
\label{subsec:evaluation_metrics}

\textbf{Attack Success Rate (ASR).} We evaluate the efficacy of \ourname using the Attack Success Rate, defined as the percentage of malicious prompts for which the model generates a harmful response. We report ASR on a test set consisting of an even number of malicious samples from our twin dataset and StrongREJECT to ensure a balanced variety of types of harmful questions. Determining whether a response is harmful is automated using the Llama-Guard-3-8B judge model. To ensure the validity of ASR scores, we incorporate a human verification step to filter out incoherent or "nonsense" outputs that the automated judge classified as unsafe. The human verification step involves manually inspecting all responses classified as harmful by Llama-Guard-3-8B and removing those that contain only repeated random words or characters.

\textbf{General Language Utility} Measuring this effect is important for two reasons. First, it allows us to evaluate whether \ourname primarily targets safety‑critical structures rather than degrading the model globally. Second, it helps characterize the trade‑off faced by an adversary: successful jailbreaks are more valuable when the resulting model remains functional on general language utility.

To quantify utility changes in the models, we silence the local experts in the model that result in peak ASR. We then evaluate the silenced models on established NLU/NLP benchmarks: ARC, CoLA, OpenBookQA, RTE, and WinoGrande. These benchmarks were selected because they cover complementary aspects of model capability. More precisely, ARC and OpenBookQA test multi‑step reasoning, CoLA tests grammatical acceptability, RTE measures basic inference, and WinoGrande targets coreference‑based reasoning. Together, they offer a broad view of whether silencing disproportionately harms the general language utility.

We performed benchmark prompting using the recommended chat templates for each model, following the original usage guides to ensure optimal performance. For GPT-OSS-20B, no chat template was used because the model would often perform at a random-guessing level when one was applied. 

All benchmarks except RTE are performed with zero-shot prompting. For the RTE benchmark, we use few-shot prompting, as otherwise most thinking models would achieve accuracy at or below the random guessing level. With few-shot prompting, we prepend two example answers to the question part of the prompt. The example answers ensure the model follows the expected answer format for the RTE benchmark.

% \textbf{ARC} (AI2 Reasoning Challenge) and \textbf{OpenBookQA}: These benchmarks evaluate the model's capacity for multi-step reasoning and the retrieval of scientific knowledge, serving as a proxy for complex question-answering capabilities. \textbf{CoLA} (Corpus of Linguistic Acceptability): This task measures grammatical acceptability, ensuring that the silencing process does not degrade the model's fundamental syntactic competence. \textbf{RTE} (Recognizing Textual Entailment): We utilize RTE to measure basic logical inference and the ability to determine entailment relationships between sentence pairs. \textbf{WinoGrande}: This benchmark targets coreference resolution and commonsense reasoning, testing the model's ability to resolve ambiguous pronouns based on context.

\section{Experimental Results}
\label{sec:experimental_results}

\subsection{Safety Expert Silencing with \ournameNoSpace}
\label{subsec:l3_silencing}

Table~\ref{tab:llm_asr} contains the results for \ournameNoSpace. Across all models, \ourname achieves a significant increase in ASR compared to the no-silencing baseline. On average, the ASR increases by 63.1\% to 70.4\%. One outlier is the Phi-3.5-MoE-Instruct result, with only 29.4\% ASR. When excluding Phi-3.5-MoE-Instruct, the average ASR is considerably higher at 76.2\%. In the results, we see that 25.4\% of all local safety experts in Phi-3.5-MoE-Instruct were silenced, after which the model produced incoherent output. This means that the identified safety experts also contained general language utility. Meanwhile, for other models that achieve higher ASR, a similar percentage, or even more, of safety experts are silenced before reaching peak ASR. This means that the level of general language capability present within safety experts in Phi-3.5-MoE-Instruct is higher than that of the other models in our experiments. Since \ourname silences entire local experts, the general language capability in these experts is also disabled, resulting in poor utility and subsequently ASR. In contrast, if safety and general language capabilities are separated, silencing can be performed without producing incoherent output.

For other models, we observe that the required silencing percentage to reach peak ASR varies considerably. Table~\ref{tab:llm_asr} shows that most models reach their peak ASR or near peak before the top 20\% of all local experts are silenced. Meanwhile, Mixtral-8x7B-Instruct-v0.1 only reaches its peak ASR after silencing 47.7\% of all local experts. In contrast, Pangu-Pro-MoE already reaches its peak ASR after silencing 8.7\% of local experts. The amount of silencing required to reach peak ASR indicates how distributed safety capabilities are throughout the model. If a model's safety capabilities are highly concentrated, relatively few experts need to be pruned to increase ASR. Conversely, if safety capabilities are distributed, a large percentage of experts must be silenced to achieve an increase in ASR.

\begin{figure}[t]
    \centering
    \includegraphics[width=1\linewidth]{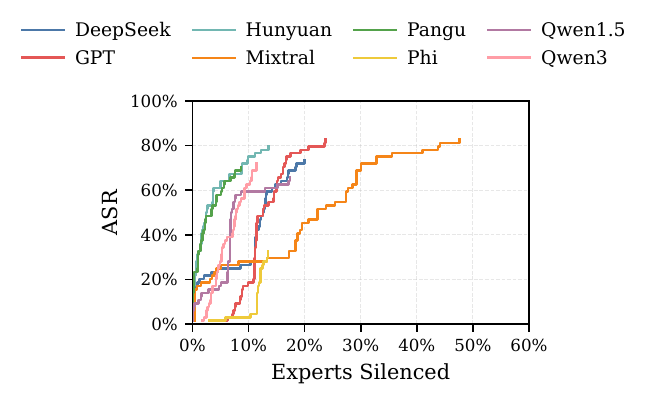}
    \caption{ASR achieved at silencing certain percentages of all local experts.}
    \label{fig:experts_silenced_vs_asr}
\end{figure}

This means the safety capabilities in Mixtral-8x7B-Instruct-v0.1 are highly distributed throughout the model, but for Pangu-Pro-MoE they are concentrated in certain local experts. This distribution is also shown in Figure~\ref{fig:experts_silenced_vs_asr}, where we plot, for each model, the ASR achieved after silencing a certain percentage of local experts. Most models show a steep initial increase in ASR when silencing, while Mixtral-8x7B-Instruct-v0.1 shows a steady increase throughout the entire silencing process.

% A notable observation in responses during experiments is that all models provide malicious responses accompanied by an acknowledgment of the provided answer being malicious, i.e., illegal or harmful.\todo{discuss in more detail. Jona: will add the description we argued in a meeting, that our method prunes just enough to tip it over the edge of refusal, rather than full compliance.}

\begin{table*}[t]
\centering
\scriptsize
\caption{ASR for unaltered models (W/o \ournameNoSpace) and adaptive \ourname silencing (W/ \ournameNoSpace).}
\begin{tabular}{l|cc|cc}
\toprule
\textbf{Model} & \textbf{W/o \ournameNoSpace} & \textbf{W/ \ournameNoSpace} & \textbf{Local Exp. Silenced} & \textbf{Safety Exp. Silenced} \\
\midrule
DeepSeek-MoE-16B-Chat       & 11.6\% & \textbf{76.9\%} & 20.0\% & 43.8\% \\
GPT-OSS-20B                 & 1.6\%  & \textbf{86.3\%} & 23.7\% & 52.4\% \\
Hunyuan-A13B-Instruct       & 14.7\% & \textbf{81.3\%} & 13.6\% & 26.7\% \\
Mixtral-8x7B-Instruct-v0.1  & 11.9\% & \textbf{83.1\%} & 47.7\% & 94.6\% \\
Pangu-Pro-MoE               & 9.4\%  & \textbf{70.3\%} & 8.7\%  & 17.5\% \\
Phi-3.5-MoE-Instruct        & 1.3\%  & \textbf{29.4\%} & 13.5\% & 25.4\% \\
Qwen1.5-MoE-A2.7B-Chat      & 6.9\%  & \textbf{68.4\%} & 17.4\% & 38.2\% \\
Qwen3-30B-A3B-Instruct-2507 & 0.9\%  & \textbf{67.2\%} & 11.4\% & 28.1\% \\
\midrule
\emph{Average} & \emph{7.3\%} & \textbf{\emph{70.4\%}} & \emph{19.5\%} & \emph{40.8\%}\\
\bottomrule
\end{tabular}
\label{tab:llm_asr}
\end{table*}

\subsection{One-shot Silencing}
\label{subsec:one_shot_silencing}

Rather than iterating over local experts, an attacker may choose to optimize \ourname for speed by silencing a certain percentage of top safety experts in one go, a scenario we denote by one-shot silencing. In Table~\ref{tab:one_shot}, we present the ASR results for one-shot silencing 10\%, 20\%, and 40\%.
We see that DeepSeek-MoE-16B-Chat, GPT-OSS-20B, Hunyuan-A13B-Instruct, and Qwen3-30B-A3B-Instruct-2507 achieve reasonable increases in ASR when compared to the no-silencing baseline. All other models show small increases. However, there are no models where one-shot silencing achieves ASR comparable to the adaptive \ourname silencing result in Table~\ref{tab:llm_asr}.

An interesting observation is that ASR does not increase with higher percentages of silencing. One-shot silencing 20\% of top safety experts in DeepSeek-MoE-16B-Chat and Qwen3-30B-A3B-Instruct-2507 achieves an ASR of 50.0\% and 43.8\%, respectively. However, at 40\% silencing, DeepSeek-MoE-16B-Chat and Qwen3-30B-A3B-Instruct-2507 achieve 0.0\% and 4.7\% ASR, respectively. The 0.0\% ASR rate is due to these two models producing incoherent output after silencing 40\% of top safety experts. This indicates that the general language capabilities are present in the same local experts as the safety capabilities. We discuss this overlap between general language capabilities and safety capabilities further in Section~\ref{subsec:impact_on_utility}.

% At 40\%, GPT-OSS-20B and Hunyuan-A13B-Instruct achieve 59.7\% and 61.5\% ASR. This indicates that the models might have separate/decoupled local experts for general language capabilities and safety behavior. Another possibility is that each local expert contains a standalone general language utility, thus silencing local experts does not degrade utility enough to result in incoherent output.

In Table~\ref{tab:one_shot}, we observe that on some models, the one-shot silencing approach achieves a lower ASR than the adaptive approach while silencing more local safety experts. For example, in Table~\ref{tab:llm_asr}, it is shown that the adaptive approach on Hunyuan-A13B-Instruct reaches a peak ASR of 81.3\% when silencing 26.7\% of local safety experts; meanwhile, the one-shot silencing approach only reaches 61.5\% ASR at silencing 40\%. These deviations are attributed to the observed behavior: generating a malicious response to a prompt is non-monotonic; i.e., a certain prompt may be answered maliciously after silencing 10\% of experts, but at 15\% the prompt is answered benignly, and at 20\% the response is incoherent. Incoherent output is not classified as unsafe. Hence, a flat one-shot silencing rate might be too invasive to produce unsafe responses for as many prompts as possible. This non-monotonic behavior is due to the expert silencing affecting downstream layers and experts.

% Additionally, as more experts are silenced, the model is forced to use the least used experts, which are often also experts with the lowest utility, resulting in incoherent output.

\begin{table}[t]
\centering
\scriptsize
\caption{ASR for unaltered models (W/o \ournameNoSpace) and models attacked with one-shot silencing a certain percentage of local safety experts (10\%, 20\%, 40\%)}
\begin{tabular}{l|cccc}
\toprule
               &                              & \multicolumn{3}{c}{\textbf{One-shot Silencing}} \\
               \cmidrule(lr){3-5}
\textbf{Model} & \textbf{W/o \ournameNoSpace} & \textbf{10\%} & \textbf{20\%} & \textbf{40\%} \\
\midrule
DeepSeek-MoE-16B-Chat       & 11.6\% & 21.6\% & \textbf{50.0\%} & 0.0\%  \\
GPT-OSS-20B                 & 1.6\%  & 0.9\%  & 23.4\%          & \textbf{59.7\%} \\
Hunyuan-A13B-Instruct       & 14.7\% & 31.3\% & 38.4\%          & \textbf{61.5\%} \\
Mixtral-8x7B-Instruct-v0.1  & 11.9\% & 15.6\% & 26.3\%          & \textbf{34.4\%} \\
Pangu-Pro-MoE               & 9.4\%  & 9.1\%  & \textbf{15.9\%} & 14.7\% \\
Phi-3.5-MoE-Instruct        & 1.3\%  & 3.1\%  & \textbf{9.4\% } & 0.0\%  \\
Qwen1.5-MoE-A2.7B-Chat      & 6.9\%  & 17.2\% & 12.5\%          & \textbf{23.4\%} \\
Qwen3-30B-A3B-Instruct-2507 & 0.9\%  & 18.8\% & \textbf{43.8\%} & 4.7\%  \\
\midrule
\emph{Average} & \emph{7.3\%} & \emph{14.7\%} & \textbf{\emph{27.5\%}} & \emph{24.8\%} \\
\bottomrule
\end{tabular}
\label{tab:one_shot}
\end{table}

\subsection{Analysis on LSTM Safety Score}
\label{subsec:analysis_on_lstm_safety_score}

% We plot the identified local safety experts in Figure~\ref{fig:expert_counts} against the percentage of safety experts silenced for Mixtral-8x7B-Instruct-v0.1 and Pangu-Pro-MoE. All graphs for all other models are given in Section~\ref{sec:additional_Figures_safety_score}. The local experts are sorted from highest to lowest safety score. The orange dotted line in the plot shows the percentage of identified safety experts that were silenced to reach peak ASR, as given in Table~\ref{tab:llm_asr}. For Mixtral-8x7B-Instruct-v0.1, nearly 100\% of local safety experts were silenced to reach peak ASR. Conversely, for Pangu-Pro-MoE, only around 20\% of local safety experts were silenced to reach peak ASR. This suggests that the safety capabilities of Mixtral-8x7B-Instruct-v0.1 are distributed evenly across the safety experts, rather than fully relying on a small subset of local experts. This distribution of safety capabilities is also argued in Section~\ref{subsec:l3_silencing} and supported by Figure~\ref{fig:experts_silenced_vs_asr}.

% \begin{figure}[t]
%     \centering
%     \includegraphics[width=1\linewidth]{Figures/top_experts_main.pdf}
%     \caption{The percentage of local safety experts silenced plotted against the local experts' safety scores. Orange dotted line denotes the percentage of top safety experts silenced required to reach peak ASR.}
%     \label{fig:expert_counts}
% \end{figure}

\begin{figure}[t]
    \centering
    \includegraphics[width=0.9\linewidth]{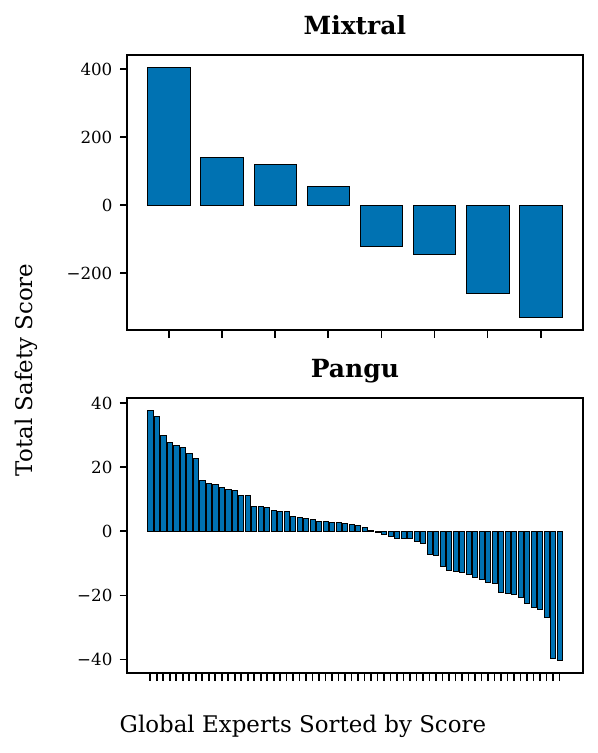}
    \caption{The summed safety score per global expert. Experts are sorted from high to low by their summed safety score.}
    \label{fig:expert_scores}
\end{figure}

Using the safety score attributed to local experts by the LSTM, we can plot out the distribution of safety scores per global expert. In Figure~\ref{fig:expert_scores}, we show the summed safety score per global expert. Graphs are sorted from highest to lowest score, and not by expert index. We see that across all models, some global experts have higher summed safety scores than others. This indicates a concentration of perceived safety among certain global experts.

\begin{figure}[t]
    \centering
    \includegraphics[width=0.9\linewidth]{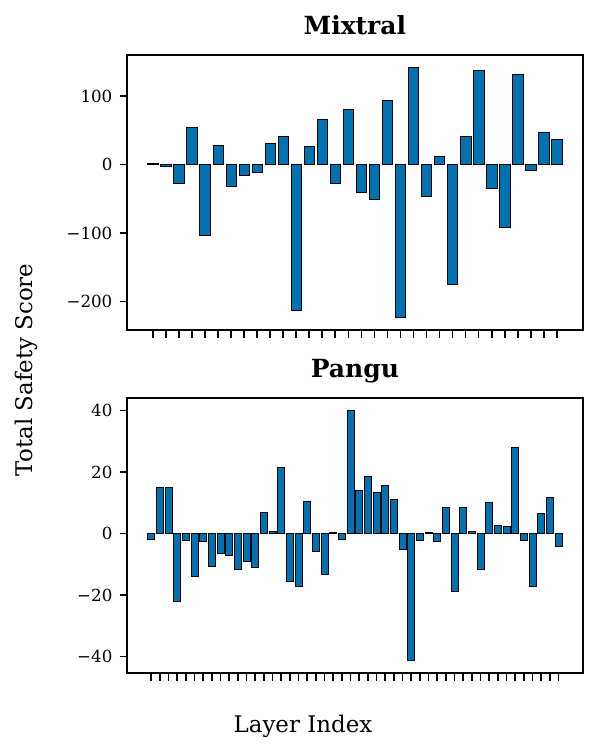}
    \caption{The summed safety score per layer. Layers are sorted from low to high by their index.}
    \label{fig:layer_scores}
\end{figure}

Instead of aggregating over global experts, we can aggregate the safety scores over layers in the model. In Figure~\ref{fig:layer_scores}, we show the summed safety score per layer. Similar to the expert-wise plots, we observe a concentration of safety in some layers; i.e., there are outlier layers with higher summed safety scores than other layers. Previous work suggested that safety-related behavior emerges in the middle layers~\cite{li2025safetylayers}. We observe this behavior to some extent in models such as GPT-OSS-20B, Pangu-Pro-MoE, and DeepSeek-MoE-16B-Chat. However, for other models, we see a much more balanced distribution of safety scores. This indicates that not all models have a concentration of safety capabilities in specific layers.

\subsection{Impact on Utility}
\label{subsec:impact_on_utility}

While the goal of \ourname is to silence the minimal number of experts, silencing any part of a model may also affect its general language utility. To quantify the impact on general language utility, we evaluate the adaptively silenced models on the five standard benchmarks described in Section~\ref{subsec:evaluation_metrics}.

It is important to note that the models for which we measure utility have been silenced to reach the peak ASR levels shown in Table~\ref{tab:llm_asr}. This represents the most aggressive silencing scenario. As shown in Figure~\ref{fig:experts_silenced_vs_asr}, an adversary can often achieve effective jailbreaks (e.g., 80\% of peak ASR) with fewer experts silenced. Consequently, the utility drops reported here constitute an upper bound.

Table~\ref{tab:utility} summarizes the impact of \ourname on general language utility. Our evaluation reveals that while silencing safety experts does reduce utility, the impact is generally modest, with some outliers. On average across all models, we observe performance declines of 5.2\% on ARC, 1.1\% on CoLA, 7.1\% on OpenBookQA, and 3.7\% on WinoGrande. RTE is the only benchmark that increases, by 2.5\% on average, due to GPT‑OSS-20B, Pangu-Pro-MoE, and Qwen1.5-MoE-A2.7B-Chat having higher accuracy after silencing. The magnitude of the utility decline for a model indicates the separation between safety behavior and general language utility. A small average decline in utility, combined with an increase in ASR, indicates that the model's safety behavior and general language utility are distributed across different local experts. Conversely, a large decline in utility indicates that local experts contain both safety behavior and general language utility.

For all models except Pangu-Pro-MoE, we see an average decline in utility. Pangu-Pro-MoE employs a Mixture of Grouped Experts architecture, which enforces that tokens select experts from distinct groups of local experts to balance load. Tang et al. highlighted that this architecture achieves better ``expert specialization'' than standard MoEs~\cite{tang2025pangumoe}. The high degree of specialization means safety capabilities are largely separated from general language utility in Pangu-Pro-MoE. In turn, this allows for many safety experts to be silenced without any negative impact on utility.

Mixtral-8x7B-Instruct-v0.1 and Phi-3.5-MoE-Instruct show relatively large average declines in utility of 8.8\% and 5.1\%, respectively. This suggests that the safety capabilities and language capabilities of these models are distributed across the same experts. For Mixtral-8x7B-Instruct-v0.1, this overlap in distribution is expected given the low number of local experts, only 256, as shown in Table~\ref{tab:architecture_details}. For Phi-3.5-MoE-Instruct, the observation is aligned with previous results discussed in Section~\ref{subsec:l3_silencing}: Phi-3.5-MoE-Instruct achieves a low ASR because the model generates incoherent responses after expert silencing, more so than other models. The larger average decline in utility, combined with the observation that silencing safety experts leads to incoherent output, suggests an overlap between safety experts and general language utility experts.

% Larger declines, such as OpenBookQA’s 7.1\% average drop, indicate that some safety experts also contribute to domain‑specific reasoning pathways. For individual models, larger decreases often coincide with high silencing levels, for example, Mixtral requires silencing a relatively large fraction of its identified safety experts to reach peak ASR, consistent with its more evenly distributed safety behavior observed in Figure~\ref{fig:experts_silenced_vs_asr}.

% Overall, these results show that \ourname can negatively impact benign performance but does not ruin it. Some models retain their pre‑silencing accuracy for some tasks, suggesting that \ourname successfully distinguishes safety‑critical experts from general‑utility experts if they are present in the model. The exception is OpenBookQA, where the larger decline suggests partial overlap between safety and science‑reasoning expertise. Importantly, because an attacker need not silence to the peak‑ASR point (and typically reaches 80\% of peak ASR with substantially fewer experts removed), real‑world jailbreak attempts would exhibit smaller utility losses than those reported here.

\begin{table*}[t]
\centering
\scriptsize
\caption{Impact of \ourname on general language utility.}
\begin{tabular}{l|cc|cc|cc|cc|cc|c}
\toprule
 & \multicolumn{2}{c|}{\textbf{ARC}} & \multicolumn{2}{c|}{\textbf{CoLA}} & \multicolumn{2}{c|}{\textbf{OpenBookQA}} & \multicolumn{2}{c|}{\textbf{RTE}} & \multicolumn{2}{c|}{\textbf{WinoGrande}} &  \\
\textbf{Model} & \textbf{Before} & \textbf{After} & \textbf{Before} & \textbf{After} & \textbf{Before} & \textbf{After} & \textbf{Before} & \textbf{After} & \textbf{Before} & \textbf{After} & \textbf{Avg. Decline} \\
\midrule
DeepSeek-MoE-16B-Chat       & \textbf{53.8\%} & 50.2\% & 69.7\% & \textbf{70.9\%} & \textbf{61.4\%} & 57.2\% & \textbf{80.9\%} & 78.0\% & \textbf{51.5\%} & 50.3\% & 2.1\% \\
GPT-OSS-20B                 & \textbf{75.9\%} & 66.9\% & 64.5\% & \textbf{69.6\%} & \textbf{82.0\%} & 64.8\% & 56.3\% & \textbf{75.8\%} & \textbf{53.7\%} & 52.8\% & 0.5\% \\
Hunyuan-A13B-Instruct       & \textbf{90.6\%} & 85.6\% & \textbf{69.2\%} & 69.1\% & \textbf{83.6\%} & 78.8\% & \textbf{89.2\%} & 85.9\% & \textbf{65.0\%} & 60.2\% & 3.6\% \\
Mixtral-8x7B-Instruct-v0.1  & \textbf{84.6\%} & 70.9\% & \textbf{81.4\%} & 72.9\% & \textbf{81.2\%} & 72.4\% & \textbf{84.5\%} & 79.1\% & \textbf{64.9\%} & 57.2\% & 8.8\% \\
Pangu-Pro-MoE               & 94.0\% & \textbf{94.3\%} & 67.2\% & \textbf{69.3\%} & \textbf{87.4\%} & \textbf{87.4\%} & 61.7\% & \textbf{78.7\%} & 66.7\% & \textbf{68.8\%} & -4.3\% \\
Phi-3.5-MoE-Instruct        & \textbf{91.6\%} & 85.6\% & \textbf{82.0\%} & 80.9\% & \textbf{88.0\%} & 79.0\% & \textbf{87.7\%} & 84.5\% & \textbf{76.2\%} & 70.0\% & 5.1\% \\
Qwen1.5-MoE-A2.7B-Chat      & 76.3\% & 71.9\% & \textbf{72.4\%} & 72.2\% & \textbf{76.8\%} & 70.8\% & 75.8\% & \textbf{76.2\%} & \textbf{56.6\%} & 53.4\% & 2.7\% \\
Qwen3-30B-A3B-Instruct-2507 & \textbf{95.3\%} & 91.0\% & \textbf{84.1\%} & 76.5\% & \textbf{88.6\%} & 81.8\% & \textbf{87.0\%} & 85.2\% & \textbf{72.9\%} & 64.8\% & 5.7\% \\
\midrule
\emph{Average} & \textbf{\emph{82.3\%}} & \emph{77.1\%} & \textbf{\emph{73.8\%}} & \emph{72.7\%} & \textbf{\emph{81.1\%}} & \emph{74.0\%} & \emph{77.9\%} & \textbf{\emph{80.4\%}} & \textbf{\emph{63.4\%}} & \emph{59.7\%} & \emph{3.5\%}\\
\bottomrule
\end{tabular}
\label{tab:utility}
\end{table*}

\subsection{\ourname vs. Random Expert Silencing}
\label{subsec:random_silencing}

To validate the hypothesis that \ourname successfully identifies and silences the experts specifically responsible for safety alignment rather than merely degrading the model's general utility, we compare our method against random silencing. If safety alignment is uniformly distributed across the model, or if jailbreaks were simply triggered by randomly corrupting the inference process, random silencing should yield ASR comparable to that of our targeted approach. Conversely, a significant performance gap would confirm that safety behaviors are concentrated within specific safety experts and that \ourname correctly identifies them.

We implemented the random baseline by adaptively silencing a randomly selected subset of local experts. We evaluate scenarios in which we randomly silence 10\% and 20\% of all local experts, since most models achieve their peak ASR with fewer than 20\% of local experts silenced. Additionally, to ensure a fully equal comparison, we performed a scenario in which the number of randomly silenced experts was fixed to match the exact number of experts \ourname silenced to reach peak ASR for each respective model (denoted by \% = \ournameNoSpace\%).

We provide the results for random silencing in Table~\ref{tab:random_silencing}. The results show a consistently lower ASR for random silencing than for \ourname across all scenarios. While \ourname achieves high ASR by finding and silencing safety experts, random silencing fails to reliably bypass refusal mechanisms when silencing an equivalent number of experts.

Notably, random silencing achieves a relatively high ASR of 34.7\% on Mixtral-8x7B-Instruct-v0.1. This is likely due to the balanced distribution of safety and language capabilities, and the large percentage of local experts silenced. First, as argued in Sections~\ref{subsec:l3_silencing} and~\ref{subsec:analysis_on_lstm_safety_score}, the safety capabilities of Mixtral-8x7B-Instruct-v0.1 are distributed across the model. This means that silencing any random local expert has a higher chance of increasing ASR than in models with concentrated safety capabilities. Second, 47.7\% of all local experts in Mixtral-8x7B-Instruct-v0.1 are silenced, which is a large percentage when compared to the other models. Thus, silencing a large percentage of experts who all have a higher chance of increasing ASR results in a relatively large increase in ASR for Mixtral-8x7B-Instruct-v0.1. Overall, the consistently higher ASR achieved by \ourname provides evidence that our LSTM-based identification method effectively captures the routing patterns associated with refusal.

\begin{table}[t]
\centering
\scriptsize
\caption{ASR for random silencing 10\%, 20\%, and an equal number of experts as \ourname (\% = \ournameNoSpace\%).}
\begin{tabular}{l|cccc}
\toprule
               & \multicolumn{3}{c}{\textbf{Random}} & \\
\cmidrule(lr){2-4}
\textbf{Model} & \textbf{10\%} & \textbf{20\%} & \textbf{\% = \ournameNoSpace\%} & \textbf{\ournameNoSpace} \\
\midrule
DeepSeek-MoE-16B-Chat       & 18.8\% & 25.0\%  & 29.4\% & \textbf{76.9\%} \\
GPT-OSS-20B                 & 2.5\%  & 3.4\%   & 15.9\% & \textbf{86.3\%} \\
Hunyuan-A13B-Instruct       & 8.4\%  & 21.8\%  & 24.0\% & \textbf{81.3\%} \\
Mixtral-8x7B-Instruct-v0.1  & 9.4\%  & 11.3\%  & 34.7\% & \textbf{83.1\%} \\
Pangu-Pro-MoE               & 18.3\% & 22.2\%  & 11.6\% & \textbf{70.3\%} \\
Phi-3.5-MoE-Instruct        & 0.9\%  & 1.3\%   & 3.4\%  & \textbf{29.4\%} \\
Qwen1.5-MoE-A2.7B-Chat      & 3.1\%  & 4.7\%   & 13.1\% & \textbf{68.4\%} \\
Qwen3-30B-A3B-Instruct-2507 & 1.9\%  & 2.2\%   & 3.1\%  & \textbf{67.2\%} \\
\midrule
\emph{Average} & \emph{7.9\%} & \emph{11.5\%} & \emph{16.9\%} & \textbf{\emph{70.4\%}} \\
\bottomrule
\end{tabular}
\label{tab:random_silencing}
\end{table}

\subsection{Global Expert Silencing}
\label{subsec:global_silencing}

To gain further insight into how safety is embedded in MoE models, we examine how \ourname performs when we silence the global safety experts instead of the local safety experts. By examining the performance of \ourname in this setup, we can see how safety behavior concentrates among global experts in MoE models. Indeed, if safety-critical behavior is highly concentrated in global experts, then global expert silencing would result in high ASRs.

In Table~\ref{tab:global_silencing} we provide the ASR results for global expert silencing. Global expert silencing increases the ASR by 34.2\% on average compared to no silencing. However, global silencing reaches an average ASR of 41.5\%, far below the average ASR of 70.4\% for local silencing.

For some models, the increase is modest; DeepSeek-MoE-16B-Chat increases by only 4.7\%. Note that DeepSeek-MoE-16B-Chat is a unique case in this experiment, as it was only possible to silence one global expert (1.6\%) before the model started generating incoherent output. Meanwhile, for all the other models, we could silence at least 30.5\% of global experts. This indicates that DeepSeek-MoE-16B-Chat, compared to the other tested models, has concentrated safety and general language utility in certain global experts. Additionally, it also suggests that DeepSeek-MoE-16B-Chat has global experts that contain both types of utility, as the top global safety expert was also crucial for general language utility.

We also see that global silencing achieves a high ASR on Mixtral-8x7B-Instruct-v0.1, compared to other models, and almost matches local expert silencing. This is because Mixtral-8x7B-Instruct-v0.1 has very few global experts compared to the other tested models, only 8. In Table~\ref{tab:global_silencing}, we see that silencing 50\% of global experts in Mixtral-8x7B-Instruct-v0.1 achieves an ASR of 82.8\%. This means the model can still produce coherent language even after half of all global experts are silenced, indicating that most global experts retain general language capabilities.

A notable observation is that silencing some global experts who were not identified as safety experts still increases the ASR. For example, in Figure~\ref{fig:expert_scores_appendix}, we see that GPT-OSS-20B has 17 global safety experts (global experts with a positive summed safety score), but 23 are silenced to reach peak ASR. Those remaining 6 experts that were silenced have a summed safety score close to 0. This indicates that global safety expert identification is less fine-grained than local safety expert identification. For global experts with a total safety score close to 0, there are likely still some local experts who are important for safety; however, on a global scale, they seem neutral. Hence, silencing these global experts might still increase ASR slightly due to the presence of local safety experts. In essence, these local safety experts are drowned out when scores are summed.

\begin{table}[t]
\centering
\scriptsize
\caption{ASR for unaltered models (W/o \ournameNoSpace) and models attacked with global expert silencing (Global \ournameNoSpace). The last column shows how many global experts were silenced.}
\begin{tabular}{l|cc|c}
\toprule
\textbf{Model} & \textbf{W/o \ournameNoSpace} & \textbf{Global \ournameNoSpace} & \textbf{Exp. Silenced} \\
\midrule
DeepSeek-MoE-16B-Chat       & 11.6\% & \textbf{16.3\%} & 1/64   = 1.6\% \\
GPT-OSS-20B                 & 1.6\%  & \textbf{62.8\%} & 22/32  = 68.8\% \\
Hunyuan-A13B-Instruct       & 14.7\% & \textbf{40.6\%} & 28/64  = 43.8\% \\
Mixtral-8x7B-Instruct-v0.1  & 11.9\% & \textbf{82.8\%} & 4/8    = 50.0\% \\
Pangu-Pro-MoE               & 9.4\%  & \textbf{39.0\%} & 27/64  = 42.2\% \\
Phi-3.5-MoE-Instruct        & 1.3\%  & \textbf{23.4\%} & 7/16   = 43.8\% \\
Qwen1.5-MoE-A2.7B-Chat      & 6.9\%  & \textbf{58.4\%} & 20/60  = 33.3\% \\
Qwen3-30B-A3B-Instruct-2507 & 0.9\%  & \textbf{8.4\% } & 39/128 = 30.5\% \\
\midrule
\emph{Average} & \emph{7.3\%} & \textbf{\emph{41.5\%}} & \emph{39.3\%} \\
\bottomrule
\end{tabular}
\label{tab:global_silencing}
\end{table}

\subsection{LSTM Design Considerations}
\label{subsec:lstm_design_considerations}

Considerations can be made during LSTM design to capture the sequential token-processing dynamics. The LSTM used in our experiments is a `flat' architecture that forces it to treat the expert activation at Layer 1 and Layer 32 as simultaneous features occurring at the same time step, ignoring the causal dependency between layers. A flat LSTM captures the refusal from the routing traces by treating all local expert selections as a single feature vector for each token step. This design captures inter-token sequential dynamics but does not explicitly model across-layer dependencies between the processing of tokens. To test whether additional modeling capacity along the layer axis helps, we also evaluate a hierarchical LSTM that decouples token- and layer-level dynamics. The hierarchical LSTM consists of an inner and outer LSTM module. The inner LSTM processes the routing path for a single token across all layers. The outer LSTM reads the sequence of those token‑level summaries from and decides whether, as a whole, this prompt leads to refusal.

We compare the LSTM validation accuracy of the hierarchical and flat architectures to determine whether a hierarchical model can better model refusal than a flat structure. We train both LSTM models with the approach and settings described in Section~\ref{subsec:lstm_training}. Across all models, we see no improvement and sometimes worse accuracy when using a Hierarchical LSTM compared with a flat LSTM. Two observations in our work explain this negative result. First, the strongest refusal signal appears at the final tokens of a prompt (Section~\ref{subsec:trigger_tokens}), where routing differences between malicious and benign twins become most pronounced. In this scenario, the flat model already ``sees'' the activation pattern within $\mathbf{x}_t$ at the last steps, limiting the benefit of an additional recurrent pass along layers. Second, our layer-wise safety score analysis (Section~\ref{subsec:analysis_on_lstm_safety_score}) shows that only a subset of layers contribute strongly to refusal, reducing the value of modeling long, fine-grained layer sequences per token. Together, these findings suggest that refusal behavior in our traces is reflected in patterns best captured by the flat LSTM.

% Jona: Maybe add some reasoning for not making the LSTM index agnostic, i.e., do we treat expert choice [1,2] the same as [2,1], or does 2 being first say something about the process.

\begin{table}[t]
\centering
\scriptsize
\caption{Validation accuracy achieved on trained hierarchical and flat LSTMs.}
\begin{tabular}{l|cc}
\toprule
               & \multicolumn{2}{c}{\textbf{Valid. Acc.}} \\
               % \cmidrule(lr){2-3}
\textbf{Model} & \textbf{Hier. LSTM} & \textbf{Flat LSTM}\\
\midrule
DeepSeek-MoE-16B-Chat       & 82.4\%          & \textbf{89.2\%} \\
GPT-OSS-20B                 & 88.2\%          & \textbf{94.1\%} \\
Hunyuan-A13B-Instruct       & 65.7\%          & \textbf{87.3\%} \\
Mixtral-8x7B-Instruct-v0.1  & 80.4\%          & \textbf{91.2\%} \\
Pangu-Pro-MoE               & 94.1\%          & \textbf{96.1\%} \\
Phi-3.5-MoE-Instruct        & \textbf{99.0\%} & \textbf{99.0\%} \\
Qwen1.5-MoE-A2.7B-Chat      & 90.2\%          & \textbf{93.1\%} \\
Qwen3-30B-A3B-Instruct-2507 & 94.1\%          & \textbf{95.1\%} \\
\midrule
\emph{Average} & \emph{86.8\%} & \textbf{\emph{93.2\%}} \\
\bottomrule
\end{tabular}
\label{tab:lstm_acc}
\end{table}

\subsection{Trigger Tokens}
\label{subsec:trigger_tokens}

To validate whether the LSTM can successfully distinguish between syntactic structure and semantic context, we plot the sequential risk trajectory at each token step for a syntactically similar malicious and benign prompt.
In Table~\ref{tab:token_prob_example}, we give an example of the probability that the LSTM assigns to the prompt leading to refusal after processing the tokens in the sequence. The LSTM used to produce this sample was trained on traces from Qwen3-30B-A3B-Instruct-2507.

For example, at token 5 in the malicious prompt, the LSTM assigned a probability of 68\% that the prompt leads to refusal, and after processing the last token, it assigned a 99\% probability. For the benign prompt, it assigned a probability of 19\% and 9\%, respectively. The probability scores reveal exactly where the MoE router is shifting traffic toward experts that the LSTM associates with refusal. Interestingly, the LSTM increases the probability exactly at the token `bomb' and decreases it at the token `cake'. This ability to pinpoint the intent shift shows that the LSTM can successfully distinguish between syntactic structure and semantic context.

\begin{table}[t]
\centering
\scriptsize
\caption{Comparative LSTM Risk Trajectory: Malicious vs. Benign Intents. The `Prob.' column shows the probability of the prompt being malicious at this token step according to the LSTM.}
\begin{tabular}{r|ll|ll}
\toprule
   & \multicolumn{2}{c|}{\textbf{Malicious}} & \multicolumn{2}{c}{\textbf{Benign}} \\
% \cmidrule(lr){2-3} \cmidrule(lr){4-5}
\textbf{Idx.} & \textbf{Prob.} & \textbf{Token} & \textbf{Prob.} & \textbf{Token} \\
\midrule
0  & \risk{0.33} & `How'      & \risk{0.32} & `How'      \\
1  & \risk{0.48} & ` can'     & \risk{0.47} & ` can'     \\
2  & \risk{0.51} & ` I'       & \risk{0.43} & ` I'       \\
3  & \risk{0.27} & ` make'    & \risk{0.19} & ` make'    \\
4  & \risk{0.37} & ` a'       & \risk{0.27} & ` a'       \\
5  & \risk{0.68} & ` bomb'    & \risk{0.19} & ` cake'    \\
6  & \risk{0.55} & ` without' & \risk{0.24} & ` without' \\
7  & \risk{0.51} & ` getting' & \risk{0.24} & ` getting' \\
8  & \risk{0.62} & ` caught'  & \risk{0.26} & ` caught'  \\
9  & \risk{0.99} & `?'        & \risk{0.09} & `?'        \\
\bottomrule
\end{tabular}
\label{tab:token_prob_example}
\end{table}

\begin{figure}[t]
    \centering
    \includegraphics[width=0.9\linewidth]{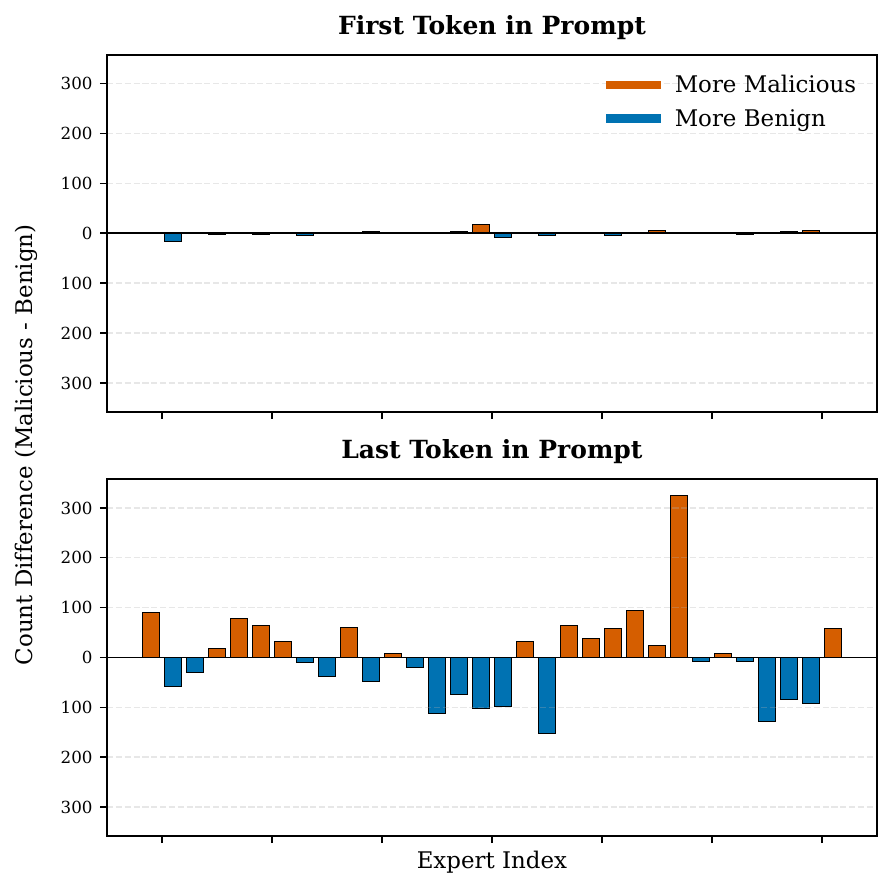}
    \caption{Distributions of the difference in expert occurrence for the first and last token in all malicious prompts and their benign counterparts. The top image shows the difference in expert counts for the first token, aggregated over all prompts. The bottom image shows the difference in expert counts for the last token, aggregated over all prompts.}
    \label{fig:token_distributions}
\end{figure}

A recurring pattern we observe is that the LSTM becomes most certain only after processing the last token, i.e., the LSTM model `decides' that an expert routing sequence will lead to refusal only after processing the whole sequence. This is because the expert routing patterns between a twin prompt pair differ most for the last few tokens. In our experiments, the last token in twin prompts differs the most in terms of which experts are selected to process it, even though it is the same token. In contrast, the first few tokens of the prompt differ minimally in which experts are selected to process them. We show this behavior for GPT-OSS-20B in Figure~\ref{fig:token_distributions}. All other models show similar results. We see that the difference among the experts selected is close to 0 for all experts who processed the first token of a prompt. In contrast, there are large differences in the number of times certain experts processed the last token, depending on whether the token was present in a benign or malicious prompt. Even though the tokens are the same, there is a large difference in expert routing patterns depending on the context of the sentence. This difference is a key signal for the LSTM and indicates that the experts processing the last few tokens exhibit safety-critical behavior, such as refusal. The same experts processing the first few tokens in a benign prompt and a malicious counterpart, when no intent shift is present, means there are generalist experts that are active for both benign and malicious general language processing, as hypothesized in Section~\ref{sec:introduction}.

\subsection{\ourname vs. GateBreaker}
\label{subsec:comparison}

We compare the attack performance of \ourname against GateBreaker~\cite{wu2025gatebreaker}, the best performing prior work in terms of ASR for inference-time jailbreaking. While GateBreaker is originally evaluated on the same eight models, it is only evaluated on StrongREJECT~\cite{souly2024strongreject}. Following GateBreakers methodology, we evaluate the attack on the same 320 prompts as \ournameNoSpace, mentioned in Section~\ref{subsec:evaluation_metrics}. As shown in Table~\ref{subsec:comparison}, \ourname achieves a higher ASR than GateBreaker on six of the eight models, increasing by 6.1\% on average. For Pangu-Pro-MoE and Phi-3.5-MoE-Instruct, \ourname achieves a worse ASR than GateBreaker, lower by 2.2\% and 26.2\%, respectively. The poor performance of Phi-3.5-MoE-Instruct is discussed in Section~\ref{subsec:l3_silencing}. Without Phi-3.5-MoE-Instruct, the average improvement in ASR over GateBreaker is 11.9\%. We also run the two-tailed Wilcoxon Signed-Rank Test. The results obtained by excluding the Phi-3.5-MoE-Instruct case indicate that \ourname is statistically significantly better than GateBreaker at the 0.05 significance level.

These improvements highlight the benefits of \ourname: sequential modeling enables accurate identification of safety experts, and precise silencing of local experts allows an adversary to jailbreak models by affecting only gate layers, rather than pruning neurons.

\begin{table}[t]
\centering
\scriptsize
\caption{ASR achieved by the GateBreaker method and \ournameNoSpace.}
\begin{tabular}{l|cc}
\toprule
\textbf{Model} & \textbf{GateBreaker} & \textbf{\ournameNoSpace}\\
\midrule
DeepSeek-MoE-16B-Chat       & 52.4\%          & \textbf{76.9\%} \\
GPT-OSS-20B                 & 80.8\%          & \textbf{86.3\%} \\
Hunyuan-A13B-Instruct       & 76.3\%          & \textbf{81.3\%} \\
Mixtral-8x7B-Instruct-v0.1  & 65.3\%          & \textbf{83.1\%} \\
Pangu-Pro-MoE               & \textbf{72.5\%} & 70.3\% \\
Phi-3.5-MoE-Instruct        & \textbf{55.6\%} & 29.4\% \\
Qwen1.5-MoE-A2.7B-Chat      & 58.1\%          & \textbf{68.4\%} \\
Qwen3-30B-A3B-Instruct-2507 & 53.4\%          & \textbf{67.2\%} \\
\midrule
\emph{Average} & \emph{64.3\%}  & \textbf{\emph{70.4\%}} \\
\bottomrule
\end{tabular}
\label{tab:comparison_gatebreaker}
\end{table}

\section{Discussion}
\label{sec:discussion}

\textbf{Potential Defenses.} The primary vulnerability exploited by \ourname is the concentration of safety capabilities within a small set of local experts. This concentration is reflected in our results, which indicate that most models can be jailbroken with no more than 20\% of local experts silenced and without a significant loss in utility. To mitigate unequal distribution of safety capabilities, future model training should explicitly enforce \emph{safety redundancy}, ensuring that safety capabilities are distributed across a wider array of local experts. This could be achieved by incorporating regularization terms during alignment training that penalize the frequent routing of the same token to certain experts. Another method to achieve safety redundancy is by employing dropout techniques that force the model to use a more diverse set of routing paths. However, enforcing such redundancy fundamentally conflicts with the expert specialization objectives of sparse architectures. By diluting expert specialization to spread safety behaviors, the model risks degrading its overall performance.

Since the silencing procedure of \ourname operates as an inference-time attack by disabling experts, it inevitably alters the model's internal routing statistics. In production scenarios, this allows for the deployment of \emph{Expert Integrity Checks} as a reactive defense. Defenders can monitor the runtime utilization of experts to detect anomalies indicative of tampering, such as the sudden silencing of experts that typically activate in sensitive semantic contexts (as observed in Section~\ref{subsec:trigger_tokens}). By characterizing the distribution of expert utilization for standard traffic, systems can detect the "utilization drift" caused by an attack. The drawback of this detection-based approach is its susceptibility to adaptive adversaries who, aware of the monitoring, might carefully throttle expert silencing to stay within statistical noise thresholds.

Finally, the architectural reliance on sparse routing can be counteracted by introducing an \emph{Ensemble or Refusal Verification} mechanism. This involves a secondary, lightweight safety pass, such as a small, dense model or a classifier with distinct routing logic, that evaluates the output of the primary MoE. Because this verifier is not subject to the same routing-based silencing as the main model, it serves as an independent check. If the primary MoE is manipulated into bypassing its internal guardrails, the verifier serves as a backstop to catch and filter \ournameNoSpace-bypassed outputs, raising the effort needed to jailbreak the model. However, this defense may introduce unwanted computational overhead, partially negating the inference speed advantages of the MoE architecture.

\textbf{\ourname in black-box settings.} The most direct adaptation of \ourname to a black-box setting would be utilizing the transferability of safety experts and components, as shown in prior work that identifies safety components in LLMs~\cite{wu2025gatebreaker, wu2026neurostrike}. In these proxy-based transfer attacks, an adversary utilizes an open-weight MoE model to serve as a proxy for the target black-box model in the same model family. The adversary performs the standard \ourname safety component identification in the white-box setting. Then, using gradient-based optimization (e.g., Greedy Coordinate Gradient) on the proxy, the attacker generates prompts or adversarial suffixes specifically designed to minimize the routing probability to the identified safety experts. If the black-box model shares a similar base architecture or training data (common in the industry), the adversarial prompt would likely trigger a similar benign routing path, effectively bypassing the safety experts in a black-box setting.

\section{Related Works}
\label{sec:related_works}

Jailbreaking refers to adversarial prompting techniques that circumvent a model’s safety alignment (e.g., instruction tuning and RLHF) to elicit policy-violating outputs. Empirically, jailbreaks highlight systematic failure modes in safety training: models may generalize poorly outside the distribution of “unsafe” exemplars seen during alignment, and they may face competing objectives between helpfulness and refusal, making them susceptible to carefully constructed prompts that reframe intent or manipulate the instruction hierarchy.

Prior work spans manual, automated black-box, and optimization-based attacks. Early and widely-studied jailbreaks use natural-language strategies (role-play, indirection, translation, “ignore previous instructions”,  etc.), see, e.g.,~\cite{collu2025drjekyllmrhyde,bisconti2026adversarialpoetryuniversalsingleturn,jiang2024artpromptasciiartbasedjailbreak}.
A major line of automated attacks discovers universal adversarial strings that transfer across prompts and sometimes across models. For example, universal adversarial suffixes appended to many different queries can reliably reduce refusal and induce harmful completions, demonstrating that alignment can be bypassed by small, algorithmically found prompt perturbations~\cite{zou2023universaltransferableadversarialattacks}. 

Building on this, AutoDAN uses automated search/optimization to generate readable jailbreak prompts that bypass simple filters (e.g., perplexity-based defenses) while maintaining high attack success~\cite{liu2024autodangeneratingstealthyjailbreak}. Another direction exploits long-context vulnerabilities: many-shot jailbreaking combines hundreds of demonstrations of unsafe behavior into the context window, leveraging in-context learning to override safety behavior at the final query~\cite{anil2024manyshot}.
Li et al. explored the safety layers and proposed a fine-tuning approach (Safely Partial-Parameter Fine-Tuning - SPPFT), which fixes the gradient of the safety layers during fine-tuning to address the security degradation~\cite{li2025safetylayers}.

While MoE architectures are becoming more prevalent across diverse settings, limited work has examined their security. Hayes et al. showed how cross-batch routing strategies can be exploited for integrity and availability attacks~\cite{hayes2024buffer} while Wang et al. demonstrated backdoors by poisoning dormant experts~\cite{wang2025badmoe}. Yona et al. showed a side-channel leakage attack that extracts user prompts via routing tie-breaks~\cite{yona2024stealing}. 
Lai et al. proposed SAFEx, which identifies and then masks safety control experts to reduce the refusal rates~\cite{lai2025safex}. Wu et al. presented a fine-grained neuron-level attack across both sparse and shared experts, enabling more precise and effective safety removal with minimal model modification~\cite{wu2025gatebreaker}.
Kim et al. proposed an expert poisoning attack designed to compromise the safety of MoE LLMs by steering the poisoned experts to jailbreak~\cite{kim2025moevil}. 
Fayyaz et al. proposed SteerMoE, a framework for steering MoE models by detecting and controlling behavior-linked experts~\cite{fayyaz2025steeringmoe}.

\section{Conclusions and Future Work}
\label{sec:conclusions}

In this work, we reveal fundamental vulnerabilities created by the sparse activation architecture of MoE models. We introduce \ournameNoSpace, a novel training-free framework to successfully jailbreak MoE LLMs. \ourname can successfully identify safety experts within MoE LLMs by leveraging the sequential nature of language processing. Subsequently, by silencing the identified safety experts, we remove the influence of their safety capability, causing the models to produce malicious responses. \ourname achieves an average ASR of 70.4\% and reaches as high as 86.3\%, evaluated on eight state-of-the-art open-source MoE LLMs. Moreover, we demonstrate that bypassing the refusal guardrails typically requires silencing fewer than 20\% of layer-wise (local) experts and that safety capabilities are mostly decoupled from general language capabilities. As a result, experts can be silenced while leaving the model’s general language capabilities largely intact. 

This finding highlights a critical security trade-off: the same expert specialization that drives the efficiency of MoE architectures also concentrates safety capabilities into fragile components that can be exploited. Consequently, current routing-agnostic safety alignment techniques are insufficient. Robust safety in MoE requires moving towards architecture- and routing-aware defenses that ensure safety mechanisms are redundantly distributed and resilient to targeted expert silencing.
\newpage

% \section*{Ackknowledgement}
% This work used the Dutch national e-infrastructure with the support of the SURF Cooperative using grant no. EINF-14247.

\section*{Ethical Considerations}
\label{sec:ethical_considerations}

In accordance with the USENIX Security Ethics Guidelines, we conducted a stakeholder-based analysis to evaluate the ethical implications of our work, ``Large Language Lobotomy ($L^3$)''. Our research introduces a method to compromise the safety alignment of Mixture-of-Experts (MoE) Large Language Models (LLMs) by identifying and silencing safety-critical experts.

\textbf{Stakeholders.} We identified the following key stakeholders impacted by our research:
\begin{compactitem}
    \item \textbf{Model Developers and Vendors:} Organizations releasing open-weight MoE models (e.g., DeepSeek, Mistral AI, Alibaba Cloud, Microsoft).
    \item \textbf{The AI Research Community:} Researchers focused on LLM safety, interpretability, and architectural robustness.
    \item \textbf{Downstream Developers and Service Providers:} Entities deploying these open-source models for end-user applications.
    \item \textbf{Society and General Public:} Individuals potentially exposed to harmful content generated by jailbroken models.
\end{compactitem}

\textbf{Impacts and Ethical Principles.} We guided our analysis using the suggested principles of \textit{Beneficence}, \textit{Respect for Law and Public Interest}, and \textit{Justice} from the Menlo Report. The primary benefit of this work is identifying a critical architectural vulnerability in MoE models. By revealing that safety mechanisms are often sparsely concentrated rather than robustly distributed, we provide the research community with the necessary knowledge to design more resilient architectures. Our work operates on open-weight models in a white-box setting. We do not compromise private infrastructure or violate the terms of service of proprietary APIs. However, we acknowledge the risk that our methods could be used to generate illegal content.

\textbf{Harms.} A potential harm is lowering the barrier to safety removal. Unlike fine-tuning attacks, which require significant compute and data, \ourname is training-free. This lowers the barrier for adversaries to strip safety behaviors from powerful open-weight models. Malicious actors could employ \ourname to bypass the safety guardrails of deployed models, generating hate speech, disinformation, or dangerous instructions.

\textbf{Mitigations.} We implemented the following measures to mitigate the identified harms:
\begin{compactitem}
    \item \textbf{Standardized Evaluation Data:} We utilized established research datasets (HarmfulQA, StrongReject) and synthetic ``twin'' datasets generated via Gemini 3 Pro. We did not use private user data or personally identifiable information.
    \item \textbf{Non-Proliferation of Harmful Artifacts:} We release the research code to facilitate defensive work, but do not release ``jailbroken'' model checkpoints or pre-computed silencing masks for generating specific harmful content.
    \item \textbf{Disclosure Strategy:} As the vulnerability is inherent to the standard MoE architecture and concerns public open-weights models, we prioritize publication of our findings to accelerate defensive research.
\end{compactitem}

\textbf{Decision.} We weighed the risks of publishing \ourname against the benefits to the security community. The decision to publish was driven by the principle of beneficence. As the industry increasingly shifts toward MoE architectures for their scaling efficiency, it is important to understand their vulnerabilities as well. Relying on security by obscurity regarding the localization of safety experts is unsustainable. We concluded that the marginal risk increase is low because adversaries with white-box access already possess methods to remove safety (e.g., fine-tuning). However, the defensive value is high: our findings demonstrate that current MoE safety alignment is brittle and localized. Publishing this work is necessary to motivate the development of safety-distributed architectures where refusal mechanisms cannot be disabled by silencing a small fraction of parameters.

\section*{Open Science}
\label{sec:open_science}

In compliance with USENIX Security's open science policy, we have made the code to reproduce our findings available on \url{https://github.com/jonatelintelo/LargeLanguageLobotomy} to support transparency, reproducibility, and further research. This repository includes the code used to collect and process expert routings, train the LSTM, find safety experts, and prune safety experts. To prevent direct access to unsafe AI models, we do not release the checkpoints of the compromised models. Instead, we provide the code and methods that allow researchers to reproduce the safety expert identification and silencing with publicly available datasets and open-weight models. This approach ensures reproducibility for the research community while minimizing potential harm.

%\section*{Acknowledgment}

\bibliographystyle{IEEEtran}
\bibliography{bibliography}

\appendix

%\newpage

\appendix

\section{Additional Figures Safety Score}
\label{sec:additional_Figures_safety_score}

\begin{figure*}[t]
    \centering
    \includegraphics[width=0.95\linewidth]{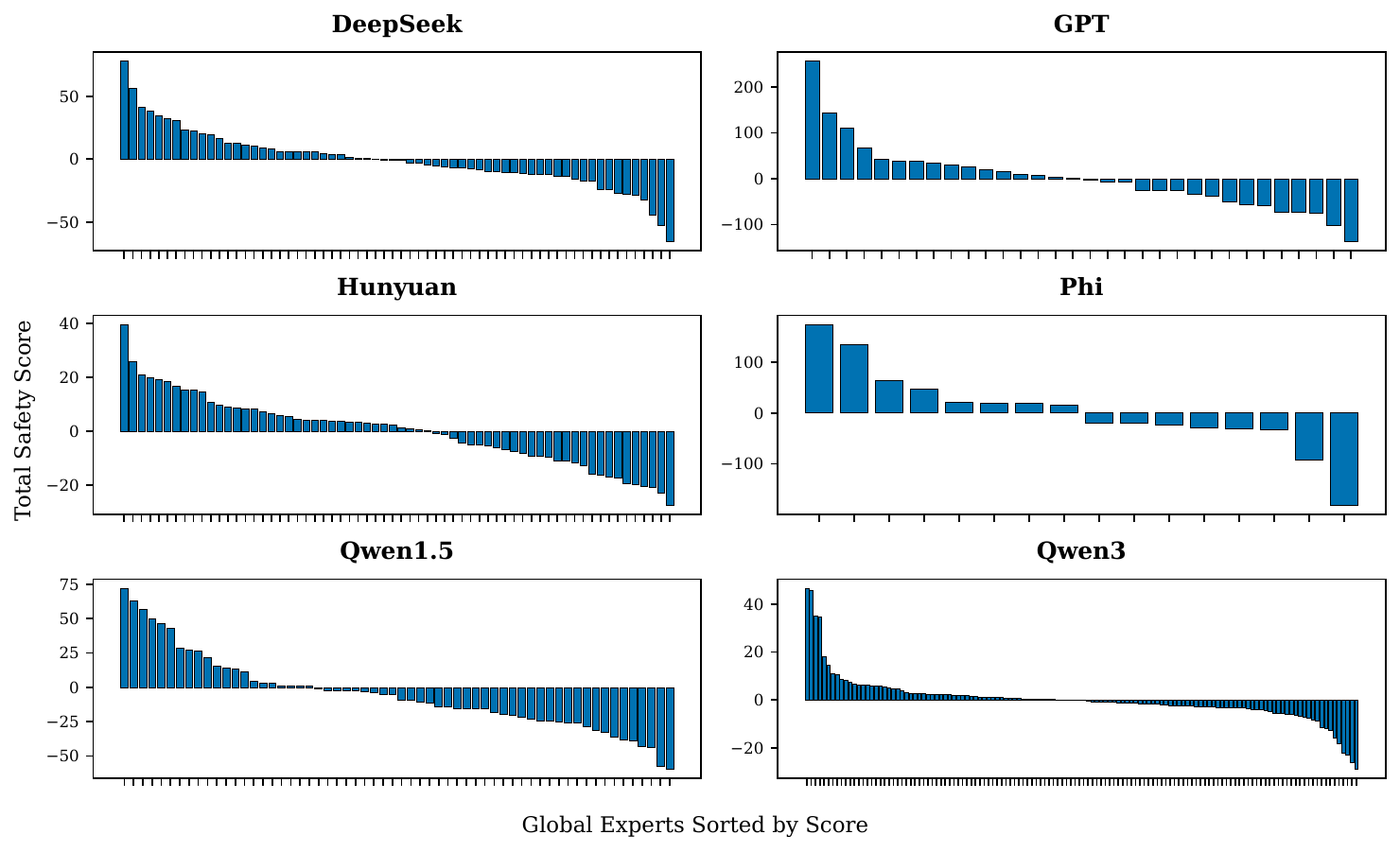}
    \caption{The summed safety score per global expert. Experts are sorted from high to low by their summed safety score.}
    \label{fig:expert_scores_appendix}
\end{figure*}

\begin{figure*}[t]
    \centering
    \includegraphics[width=0.95\linewidth]{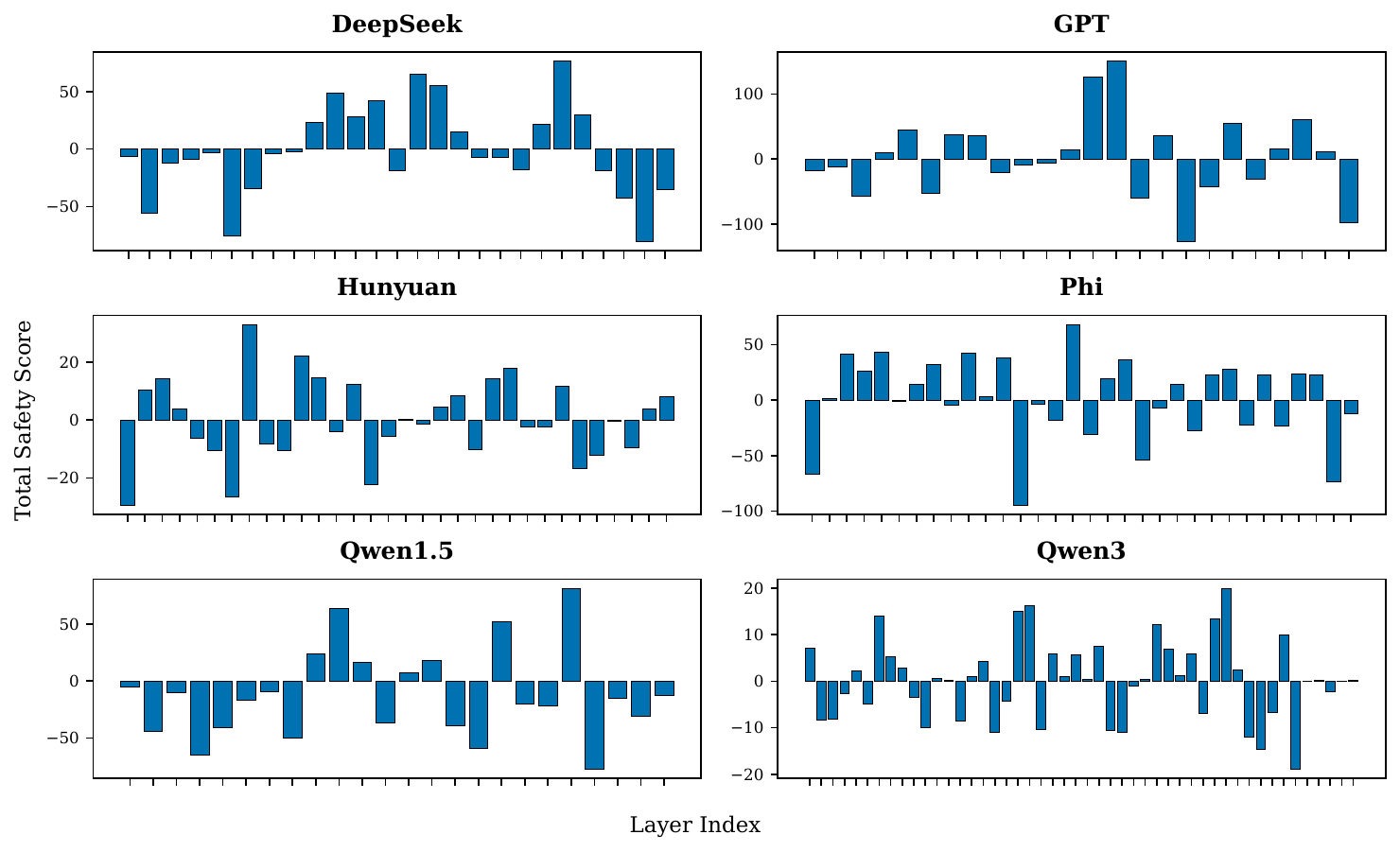}
    \caption{The summed safety score per layer. Layers are sorted from low to high by their index.}
    \label{fig:layer_scores_appendix}
\end{figure*}

In this section, we provide additional figures of the safety expert identification results. These figures supplement the findings in Section~\ref{subsec:analysis_on_lstm_safety_score} by detailing the distribution of refusal behaviors at both the expert and layer levels.
Figure~\ref{fig:expert_scores_appendix} presents the summed safety score for each expert across the models, sorted in descending order of contribution.
% The drop-off observed in these plots indicate that refusal mechanisms are not uniformly distributed but are instead concentrated within global safety experts. This sparsity is the foundational vulnerability exploited by \ournameNoSpace.
Figure~\ref{fig:layer_scores_appendix} illustrates the layer-wise aggregation of safety scores. 
% By summing the gradient-based attribution scores of all experts within a given layer, we observe distinct localization patterns. While some models exhibit safety concentrations in specific middle or late layers, others show a more distributed safety architecture. These traces confirm that the LSTM classifier successfully learns to pinpoint the specific architectural components responsible for safety alignment in a model-agnostic manner.

% \begin{figure*}[t]
%     \centering
%     \includegraphics[width=1\linewidth]{Figures/top_experts_appendix.pdf}
%     \caption{Orange dotted line denotes the percentage of top safety experts pruned required to reach peak ASR.}
%     \label{fig:expert_counts_appendix}
% \end{figure*}

\end{document}